%% file: strange.tex
\documentclass[12pt,a4paper,dvips]{article}
\usepackage{a4p}
\usepackage{cite} 
\usepackage{graphicx}
\usepackage{physics}
\usepackage{l3_titlePH,ifthen,Lep}
\usepackage{epsfig,float}
\usepackage{rotating}
\journalname{Eur. Phys. J. C}
\preprint{2006-028}
\date{July 12, 2006}
\def\ra{\rightarrow }

\def\epem{\mbox{e}^+\mbox{e}^- }

\def\wgg{W_{\gamma \gamma}}
\def\ra{\rightarrow } 
\def\epem{\mbox{e}^+\mbox{e}^- } 
\def\qqbar{\rm q \overline{q}}
 
\def\mpko{m(\rm p K^0_S)}
\def\phz{\phantom{0}}

\newlength{\capindent}
\newlength{\capwidth}
\newlength{\figwidth}
\newcommand{\icaption}[2][!*!,!]{\hspace*{\capindent}%
  \begin{minipage}{\capwidth}
    \ifthenelse{\equal{#1}{!*!,!}}%
      {\caption{#2}}%
      {\caption[#1]{#2}}
  \end{minipage}}

\begin{document}
\begin{titlepage}
\title{Study of Inclusive Strange-Baryon Production and\\ 
Search for Pentaquarks in Two-Photon Collisions at LEP}
\author{The L3 Collaboration}
\begin{abstract}
Measurements of inclusive production of the $\Lambda$, $\Xi^-$ and $\Xi^*(1530)$ 
baryons in two-photon collisions with the L3 detector at LEP are presented. 
The inclusive differential cross sections for $\Lambda$ and $\Xi^-$ are measured as 
a function of the baryon transverse momentum, $p_t$, and pseudo-rapidity, $\eta$. 
The mean number of $\Lambda$, $\Xi^-$ and $\Xi^*(1530)$ baryons per hadronic 
two-photon event is determined in the kinematic range $0.4 \GeV < p_t< 2.5 \GeV$, $|\eta| < 1.2$. 
Overall agreement with the theoretical models and Monte Carlo predictions is observed.
A search for inclusive production of the pentaquark $\theta^+(1540)$ in two-photon 
collisions through the decay $\theta^+ \ra \rm p \kos$ is also presented. No evidence 
for production of this state is found.
\end{abstract}
\submitted
\end{titlepage}

\section{Introduction}

Two-photon collisions are the main source of hadron production in high-energy $\rm \epem$ 
interactions at LEP, via the process ${\rm e}^{+} {\rm e}^{-} \rightarrow {\rm e}^{+} 
{\rm e}^{-} \gamma ^{*} \gamma ^{*}  \rightarrow  {\rm e}^{+} {\rm e}^{-}  hadrons$, for 
which the cross section is many orders of magnitude larger than the $\epem$ annihilation cross section. 
The outgoing electron and positron carry almost the full beam energy and their 
transverse momenta are usually so small that they escape undetected along the beam pipe. 
At the LEP energies considered here, the negative four-momentum squared of the photons, $Q^2$, has an 
average value of $\langle Q^2 \rangle \simeq 0.2 \GeV^2$. Therefore, the photons may be  
considered as ``quasi-real''. In the Vector Dominance Model (VDM), each virtual photon 
can fluctuate into a vector meson, thus initiating a strong interaction 
process with characteristics similar to hadron-hadron interactions. This process 
dominates in the ``soft'' interaction region, where hadrons are produced with a 
low transverse momentum, $p_t$. Hadrons with high $p_t$ are instead mainly produced 
by the QED process $\gamma \gamma \ra \qqbar$ (direct process) or by QCD processes 
originating from the partonic content of the photon (resolved processes). 

Fragmentation mechanisms can be investigated in two-photon reactions. These processes 
are described phenomenologically. In the Lund string 
model \cite{lund_string}, hadron production proceeds through the creation of 
quark-antiquark and diquark-antidiquark pairs during string fragmentation. Mesons 
and baryons are formed by colorless quark-antiquark and quark-diquark 
combinations,  respectively. An extension of this model, the ``simple popcorn mechanism'' 
\cite{popcorn}, includes the possibility of producing an additional meson 
between baryon-antibaryon pairs. The relative rate of occurrence of the baryon-meson-antibaryon 
configuration is governed by the so called ``popcorn parameter.'' Many other 
parameters must be tuned to reproduce the measured hadron production rate, such as 
the strange-quark suppression factor, the diquark-to-quark production ratio 
or the spin-1 diquark suppression factor. 

In a statistical model approach~\cite{thermotheo},
hadronisation is described with a reduced number of free parameters.
Particles are produced in a purely statistical way from a massive
colourless gas which, for hadron production in two-photon collisions, is completely specified by two
parameters: the
energy density (or temperature) and a strange-quark suppression factor.
In this model, the yield of different particles depends only on their
masses, spins and quantum numbers.

The L3 Collaboration has previously measured inclusive $\pi^0$, $\kos$ and
$\Lambda$ production in quasi-real two-photon collisions for a centre-of-mass energy 
of the two interacting photons, $\wgg$, greater than $5 \GeV$ \cite{pz,l3lambda}. In this Article 
this study is extended to the $\Xi^-$, $\Xi^*(1530)$ and $\Omega^-$ 
baryons\footnote{If not stated otherwise, the symbols $\Lambda$, $\Xi^-$ and $\Omega^-$ 
refer to both the $\Lambda$, $\Xi^-$ and $\Omega^-$ as well as $\overline{\Lambda}$, 
$\overline{\Xi}^+$ and $\overline{\Omega}^+$ baryons. All charge-conjugate final-states are analysed.}. 
The data sample consists of  a total integrated luminosity of 610 pb$^{-1}$, collected with the L3 detector 
\cite{L3col} at $\rm \epem$ centre-of-mass energies $\sqrt{s}=189 - 209 \GeV$, with 
a luminosity weighted average value $\langle \sqrt{s} \rangle=198 \GeV$. Inclusive strange-baryon 
production has been extensively studied in $\epem$ annihilation processes \cite{pdg}, whereas 
in two-photon collisions only inclusive $\Lambda$ production  has been previously 
measured at lower $\sqrt{s}$ by the TOPAZ Collaboration \cite{topaz}. 

Heavy-baryon detection techniques are also applied to search for pentaquark production in two-photon 
interactions, using the decay channel $\theta^+ \ra \rm p \kos$. 
Since the first observation of a narrow resonance near $1540 \MeV$ in the $\rm K^+ n$
mass spectrum \cite{leps}, interpreted as the pentaquark $\theta^+$, experimental evidence 
for and against this new state has been accumulated \cite{pentasearchold}. Second-generation 
experiments have so far not confirmed the existence of this state \cite{pentasearchnew}. 
No search in two-photon reactions has been performed yet.

\section{Monte Carlo simulation}

The process $\epem \ra \epem hadrons$ is modelled with the PYTHIA
\cite{pythia} and PHOJET \cite{phojet} event generators with twice the 
statistics of the data. In PYTHIA, each photon is classified as direct, 
VDM or resolved, leading to six classes of two-photon events. 
A smooth transition between these classes is obtained by introducing a $p_t$ 
parameter to specify the boundaries. The SaS-1D parametrization is used for 
the photon parton density function~\cite{sas1d}. 
Since both incoming photons are assumed to be real in the original program, PYTHIA 
is modified to generate a photon flux according to the equivalent photon 
approximation~\cite{budnev} with an upper $Q^2$ cut at the mass squared of the rho meson.

PHOJET is a general purpose Monte Carlo which describes hadron-hadron, photon-hadron 
and photon-photon collisions. It relies on the dual parton model combined with the
QCD-improved parton model~\cite{dpm}. The $p_t$ distribution of the 
soft partons is matched to the one predicted by QCD to ensure a continuous transition 
between hard and soft processes. The two-photon luminosity function is calculated in
the formalism of Reference~\citen{budnev}. The leading-order GRV
parametrisation is used for the photon parton density function~\cite{grv}.  

In both programs, matrix elements and hard scattering processes are calculated at 
the leading order and higher-order terms are approximated by a parton shower 
in the leading-log approximation. The fragmentation is performed within the Lund string
fragmentation scheme as implemented in JETSET \cite{pythia}, which is also used
to simulate the hadronisation process. JETSET parameters are tuned by using 
hadronic Z decays \cite{QCDtuning}. In particular, 
the strangeness-suppression factor is set to 0.3,
whereas the parameter governing the extra suppression of strange quarks in diquarks 
is fixed to 0.4 and the diquark-to-quark production ratio to 0.10. The popcorn parameter is 
set to 0.5 and a value $\alpha_S(m_{\rm Z})=0.12$ is used for the strong coupling-constant.

The following Monte Carlo generators are used to simulate the  
background processes: KK2f\cite{KK2f} for the annihilation process $\epem \rightarrow \rm \qqbar \,
(\gamma $); KORALZ \cite{KORALZ} for $\epem \rightarrow \tau^{+} \tau^{-}(\gamma )$; 
KORALW \cite{KORALW} for $\epem \rightarrow \rm{W}^{+} \rm{W}^{-}$  and  
DIAG36 \cite{DIAG36} for $\epem \ra  \epem \tau^{+} \tau^{-}$. 
The response of the L3 detector is simulated using the GEANT \cite{GEANT} 
and GHEISHA \cite{GEISHA} programs. Time-dependent detector inefficiencies, as monitored 
during each data-taking period, are included in the simulations. All simulated events 
are passed through the same reconstruction program as the data.

\section{Two-photon event selection}

Two-photon interaction events are mainly collected by track triggers 
\cite{trigtrack}, with a track $p_t$ threshold of about 150 \MeV, 
and the calorimetric energy trigger \cite{trigener}. The selection of  
${\rm e}^{+} {\rm e}^{-} \rightarrow {\rm e}^{+} {\rm e}^{-} hadrons$ 
events \cite{l3tot} is based on information from the central tracking detectors and 
the electromagnetic and hadronic calorimeters. It consists of:

\begin{itemize}

\item A multiplicity cut. To select hadronic final states, at  
least six objects must be detected, where an object can be a 
track satisfying minimal quality requirements or an isolated 
cluster in the BGO electromagnetic calorimeter of energy greater than 100 \MeV.

\item 
Energy cuts. The total energy deposited in the calorimeters
must be less than 40\% of $\sqrt{s}$ to suppress events from the 
$\epem \ra \rm q \overline{q} (\gamma)$ and $\epem \ra \tau^+ \tau^- (\gamma)$ processes.
In addition, the total energy in the electromagnetic calorimeter is required to be greater 
than $500 \MeV$ to suppress beam-gas and beam-wall interactions
and less than $50 \GeV$ to remove events from the annihilation process 
$\epem \ra \rm q \overline{q} (\gamma)$.

\item
An anti-tag condition. Events containing a cluster in the luminosity monitor with an 
electromagnetic shower shape, and energy greater than $30 \GeV$, are excluded 
from the analysis. The luminosity monitor covers the polar angular region 
31 mrad $<\theta<$ 62 mrad on both sides of the detector. In addition, 
events with an electron or positron scattered above 62 mrad are rejected by the cut 
on calorimetric energy.

\item
A mass cut. The mass of all visible particles, $W_{vis}$, must 
be greater than $5 \GeV$ to exclude the resonance region. In this calculation, the pion 
mass is attributed to tracks while isolated electromagnetic clusters are treated as massless.

\end{itemize}

About 3 million hadronic events are selected by these criteria with an 
overall efficiency of 45\% for a two-photon centre-of-mass energy, $\wgg$, greater 
than $5 \GeV$. The background level of this sample is 
less than 1\% and is mainly due to the $\epem \ra \rm q 
\overline{q}(\gamma)$ and $\epem \ra \epem \tau^+ \tau^-$ processes.
The backgrounds from beam-gas and beam-wall interactions are negligible.

\section{Strange baryon selection}

The $\Lambda$, $\Xi^-$ and $\Omega^-$ baryons are identified through the decays 
$\Lambda \ra \rm p \pi^-$, $\Xi^- \ra \Lambda \pi^-$ and 
$\Omega^- \ra \Lambda \rm K^-$. Due to their long lifetimes, the combinatorial 
background can be strongly suppressed by selecting secondary vertices which 
are clearly displaced from the interaction point. The $\Xi^*(1530)$ decays strongly 
into $\Xi^- \pi^+$ and no corresponding displaced vertex can be observed. 

The $\Xi^-$, $\Omega^-$ and $\Xi^*(1530)$ final states are reconstructed 
by combining $\Lambda$ candidates with pion or kaon candidates. The latter are defined as 
tracks formed by at least 30 hits in the central tracker out of a maximum of 62 and 
a $p_t > 100 \MeV$. The probability of the 
pion or kaon hypothesis, based on the d$E$/d$x$ measurement in the tracking 
chamber, is required to be greater than 0.01.

\subsection{$\Lambda$ selection}

The selection procedure is unchanged from our previous publication 
\cite{l3lambda}. The $\Lambda$ identification is optimized to achieve a high efficiency 
and a good background suppression by selecting secondary decay vertices satisfying 
the following conditions:

\begin{itemize}

 \item  The distance $d_\Lambda$ in the transverse plane\footnote{The transverse plane is defined as the plane 
  transverse to the beam direction.} between the secondary 
  vertex and the $\epem$ interaction point must be greater 
  than 3 mm.

 \item The angle $\alpha_\Lambda$ between the sum $p_t$ vector of 
 the two tracks and the direction in the transverse plane between the 
 $\epem$ interaction point and the secondary vertex must be less than 100 mrad. 
\end{itemize}

\begin{figure}
\begin{center}
\epsfig{file=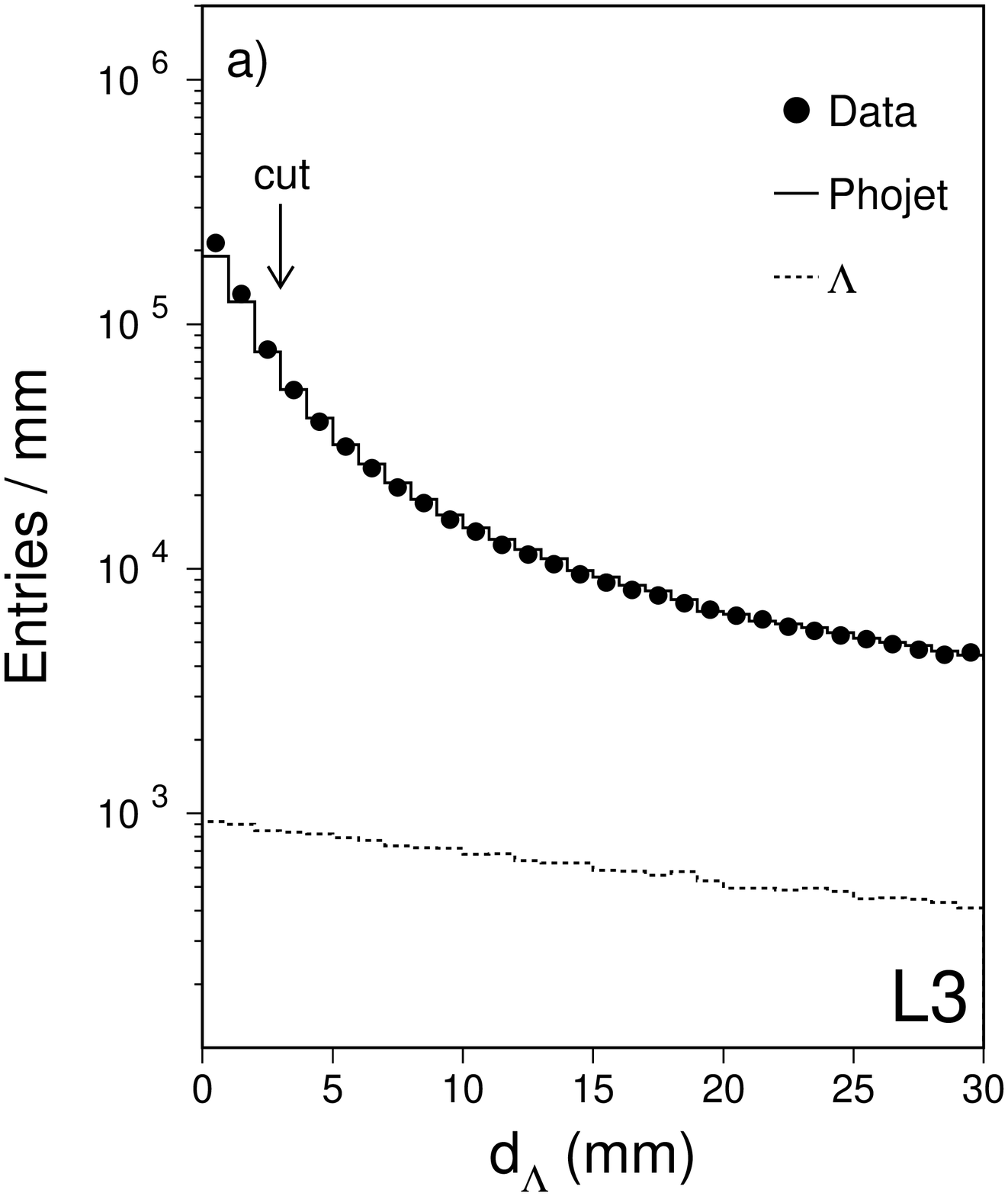,width=7.5cm} 
\epsfig{file=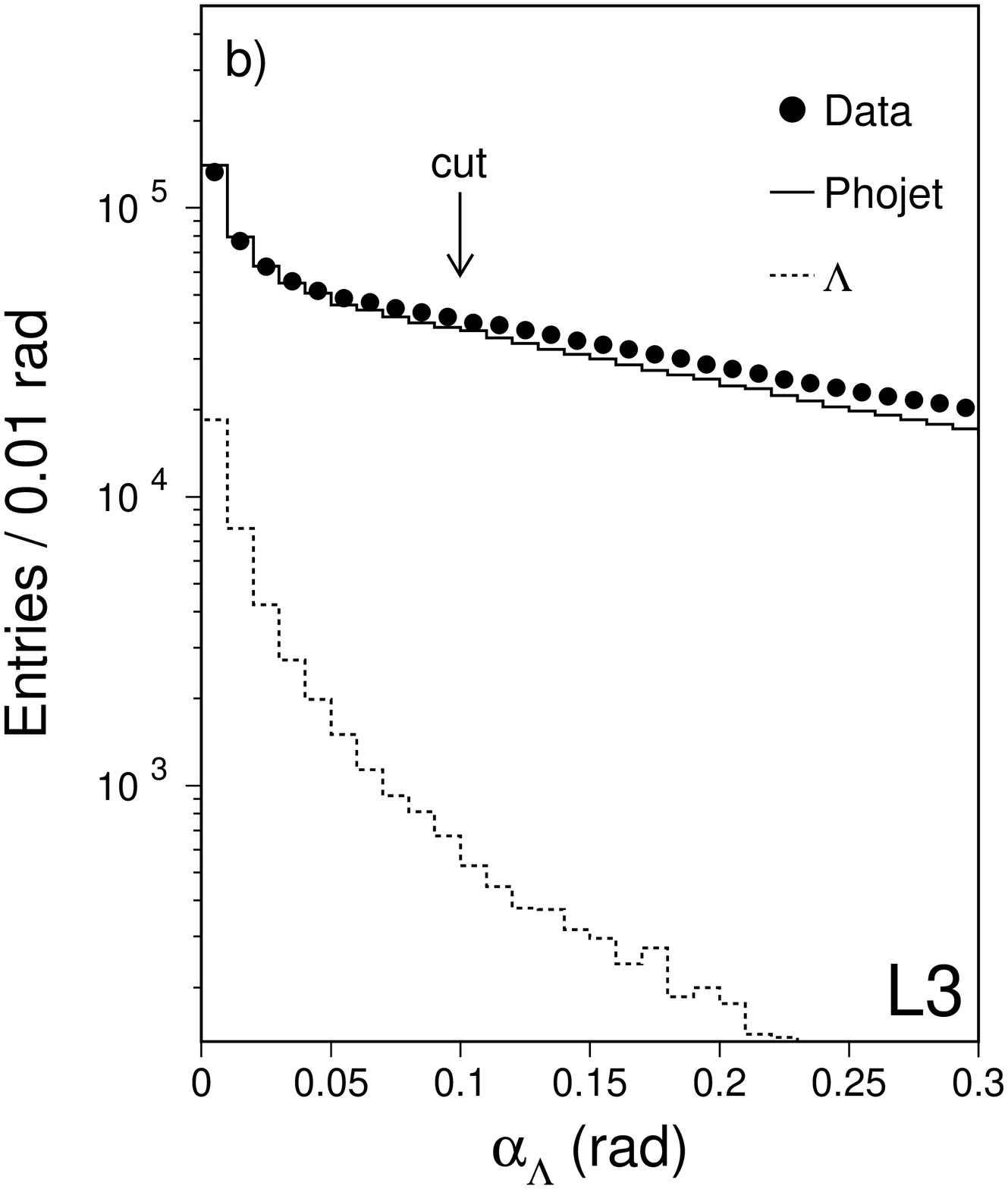,width=7.5cm}
\end{center}
\caption{Distribution of the variables used to select secondary
vertices from inclusive $\Lambda$ production: a) the distance in the transverse plane between the secondary vertex and the 
$\epem$ interaction point, $\rm d_\Lambda$, and b) the angle between the sum $p_t$ vector of 
the two tracks and the direction in the transverse plane between the interaction 
point and the secondary vertex, $\alpha_\Lambda$. All other secondary vertex selection 
criteria are applied. The predictions of the PHOJET Monte Carlo are shown as the solid  
lines and the contribution due to $\Lambda$ baryons as the dashed lines. The Monte Carlo 
distributions are normalized to the data luminosity.}
\label{lamcutvtx}
\end{figure}

The distributions of these variables are presented in Figure \ref{lamcutvtx}. Good 
agreement between data and Monte Carlo is observed. The proton is identified 
as the track with the largest momentum, an assignment shown by Monte Carlo to 
be correct more than 99\% of the time. The d$E$/d$x$ measurements in the central 
tracking chamber must be consistent with this assignment, a probability 
greater than 0.01 for both the proton and the pion candidates being required. 

\begin{sidewaysfigure}
\begin{center}
\epsfig{file=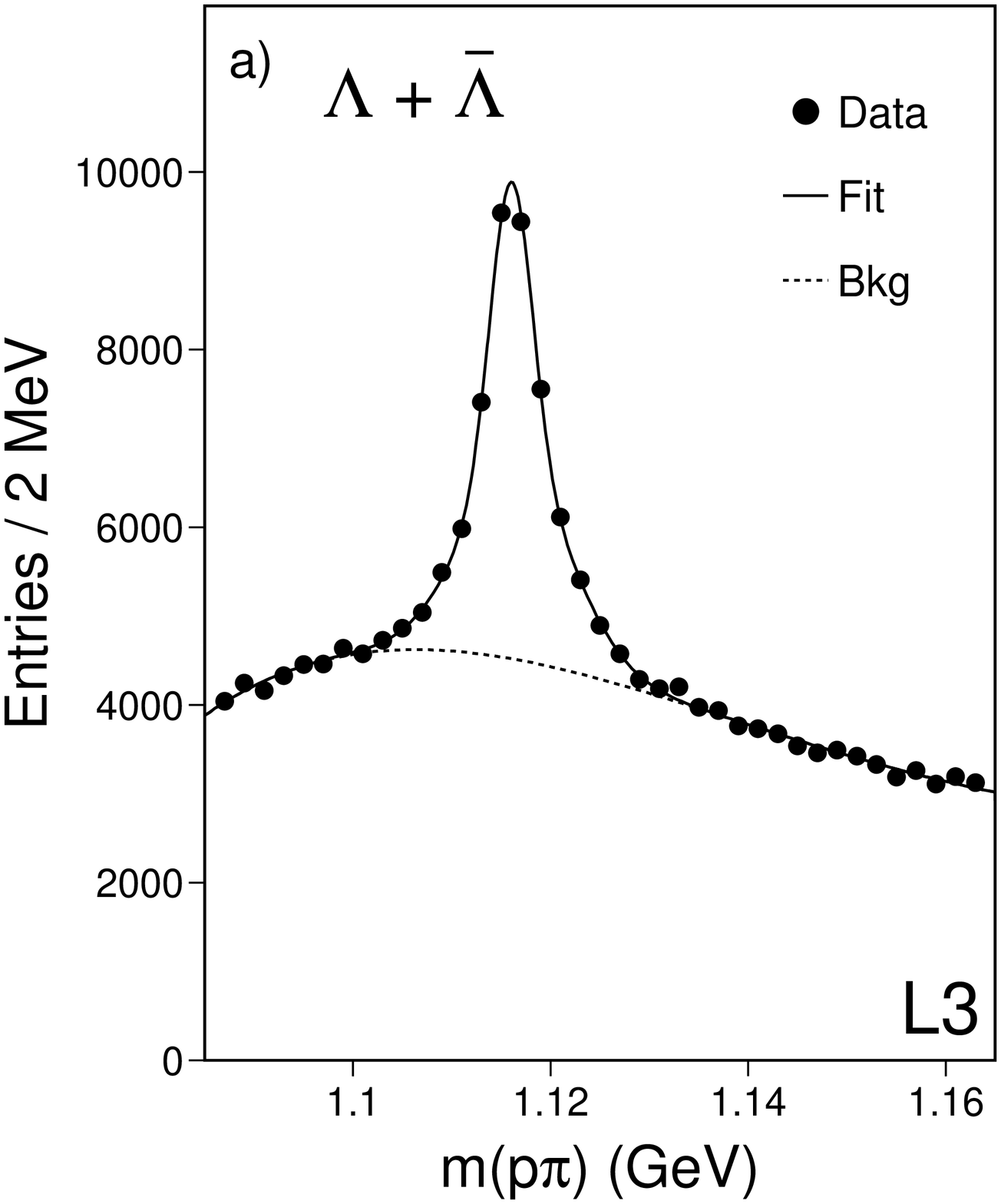,width=7.5cm}
\epsfig{file=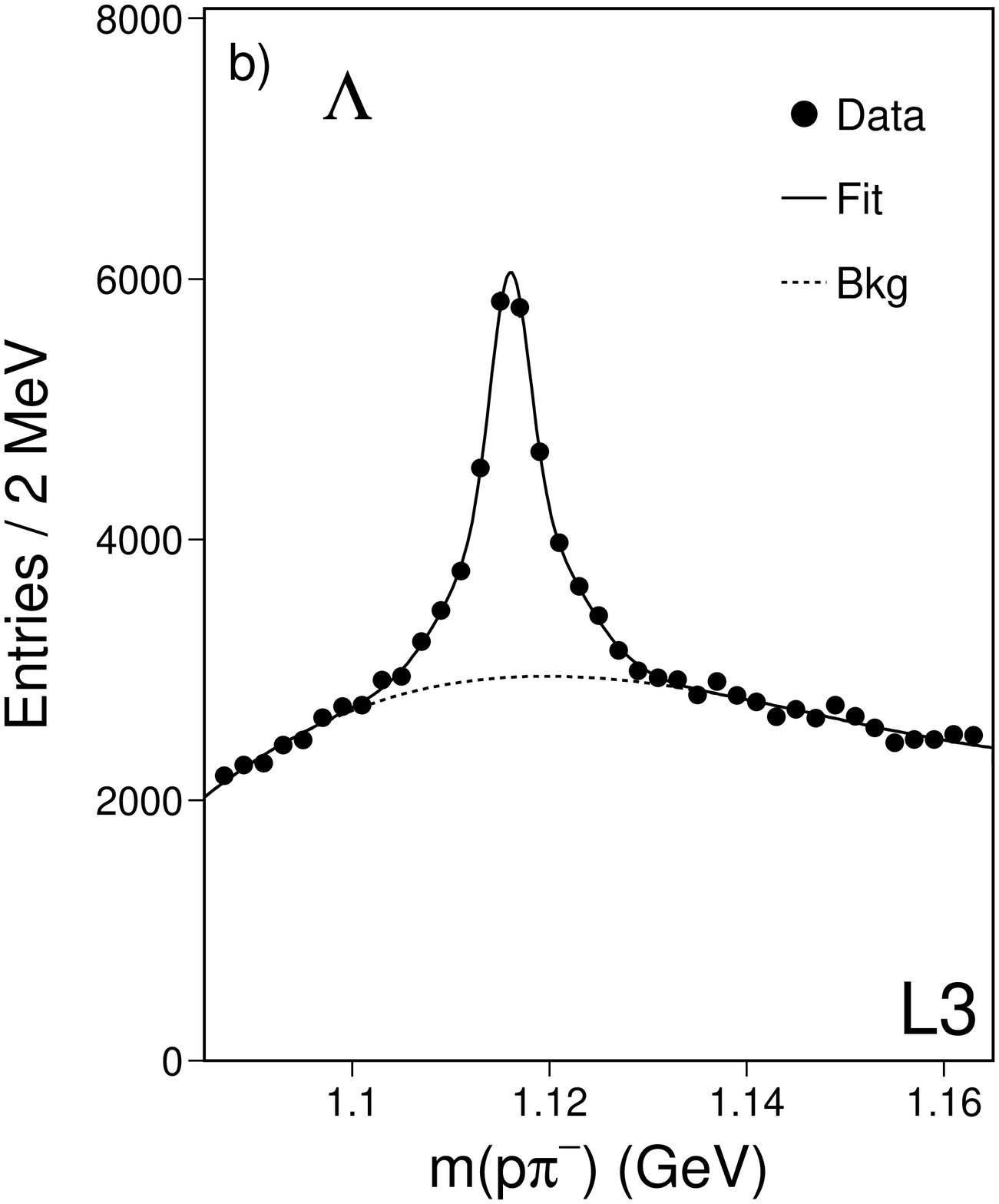,width=7.5cm}
\epsfig{file=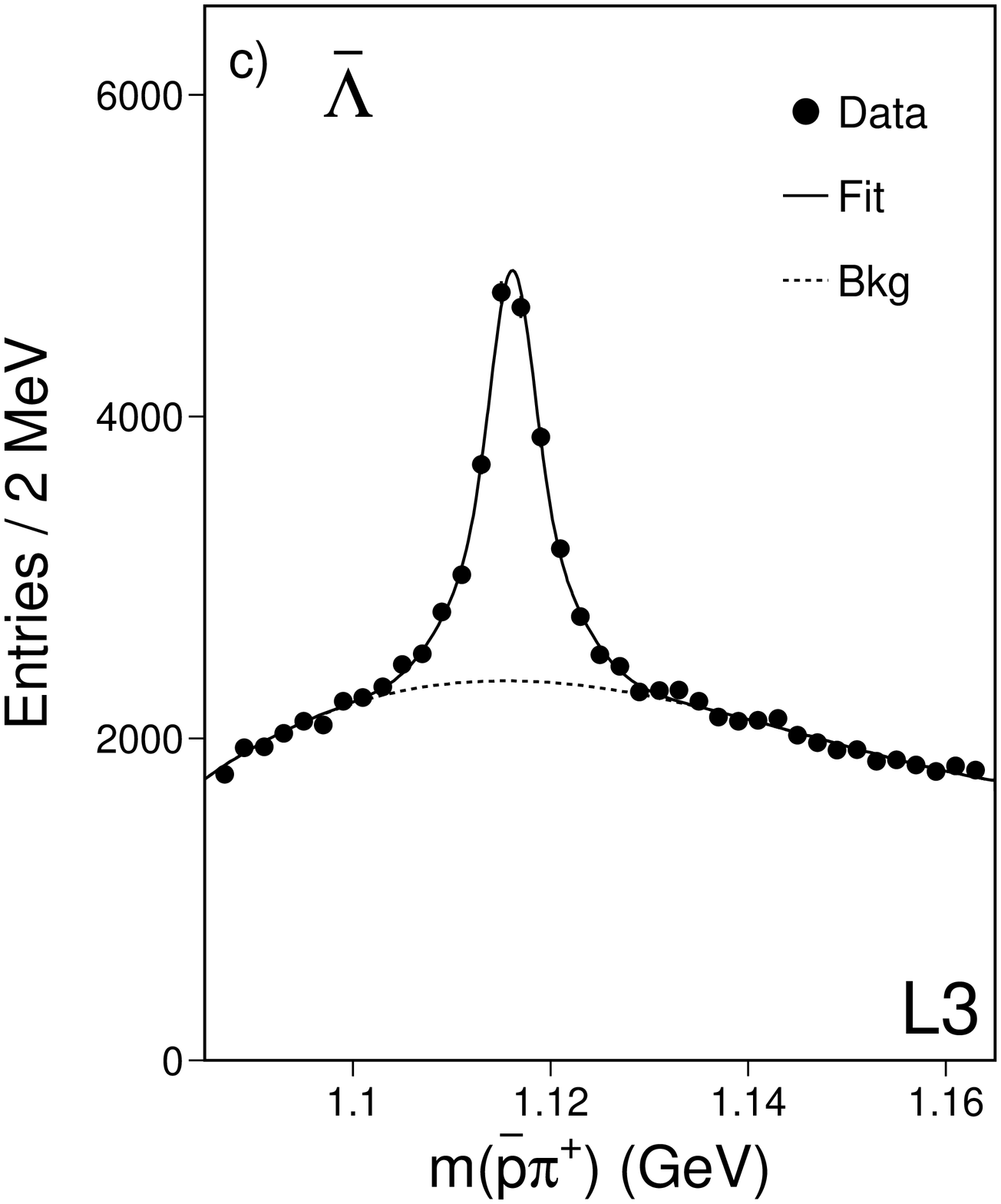,width=7.5cm}
\end{center}
\caption {The mass of the $\rm p \pi$ system for the inclusive $\Lambda$ 
selection for a) $\Lambda$ and $\overline{\Lambda}$ candidates, b) $\Lambda$ candidates and c) 
$\overline{\Lambda}$ candidates. The signal is modelled with two Gaussians 
and the background by a fourth-order Chebyshev polynomial.}
\label{lambdaplot}
\end{sidewaysfigure}

The distribution of the corresponding invariant mass of the ${\rm p} \pi$ system, 
$m({\rm p} \pi)$, is shown in Figure \ref{lambdaplot}a. A clear $\Lambda$ peak over a smooth 
background is visible, compatible with the $\Lambda$ baryon mass, 
$m_{\Lambda} = 1115.68\pm 0.01 \MeV$ \cite{pdg}. The resolution of the signal, about 
$3 \MeV$, is well reproduced by Monte Carlo simulation. The $m({\rm p} \pi)$ mass spectrum 
for the $\rm p \pi^-$ and $\rm \overline{p} \pi^+$ combinations are shown in Figure 
\ref{lambdaplot}b and \ref{lambdaplot}c, respectively.
The numbers of $\Lambda$ baryons for different $p_t$ and $|\eta|$ bins are given in 
Tables \ref{tabcrosspt} and \ref{tabcrosseta}. They are determined by a fit in which the signal 
is modelled by a Gaussian and the background by a fourth-order Chebyshev polynomial. 
Consistent values of the fitted mass and width are found for the different $p_t$ 
and $|\eta|$ bins.

\subsection{$\Xi^-$ selection}

\begin{sidewaysfigure}
\begin{center}
\epsfig{file=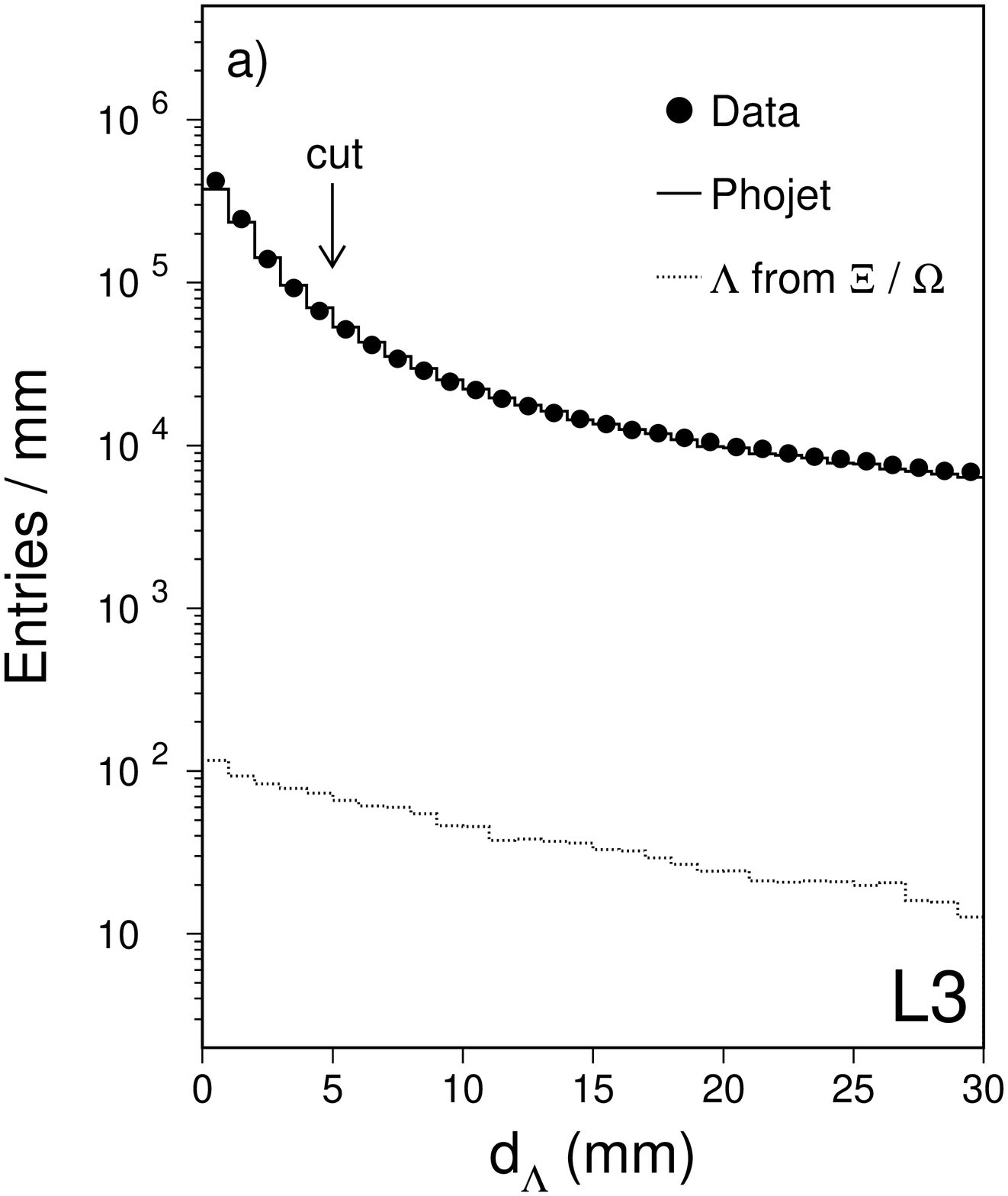,width=7.5cm} 
\epsfig{file=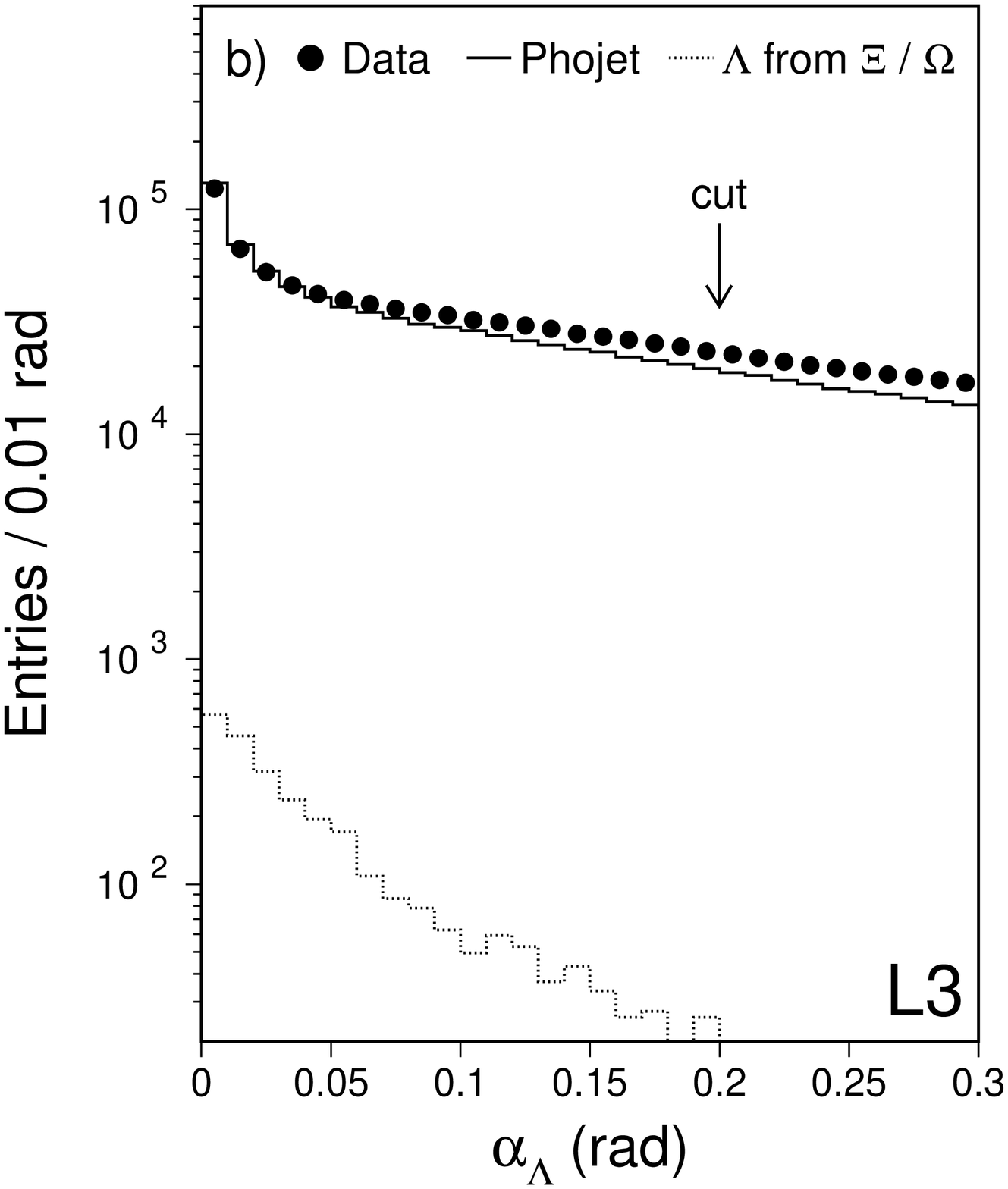,width=7.5cm}
\epsfig{file=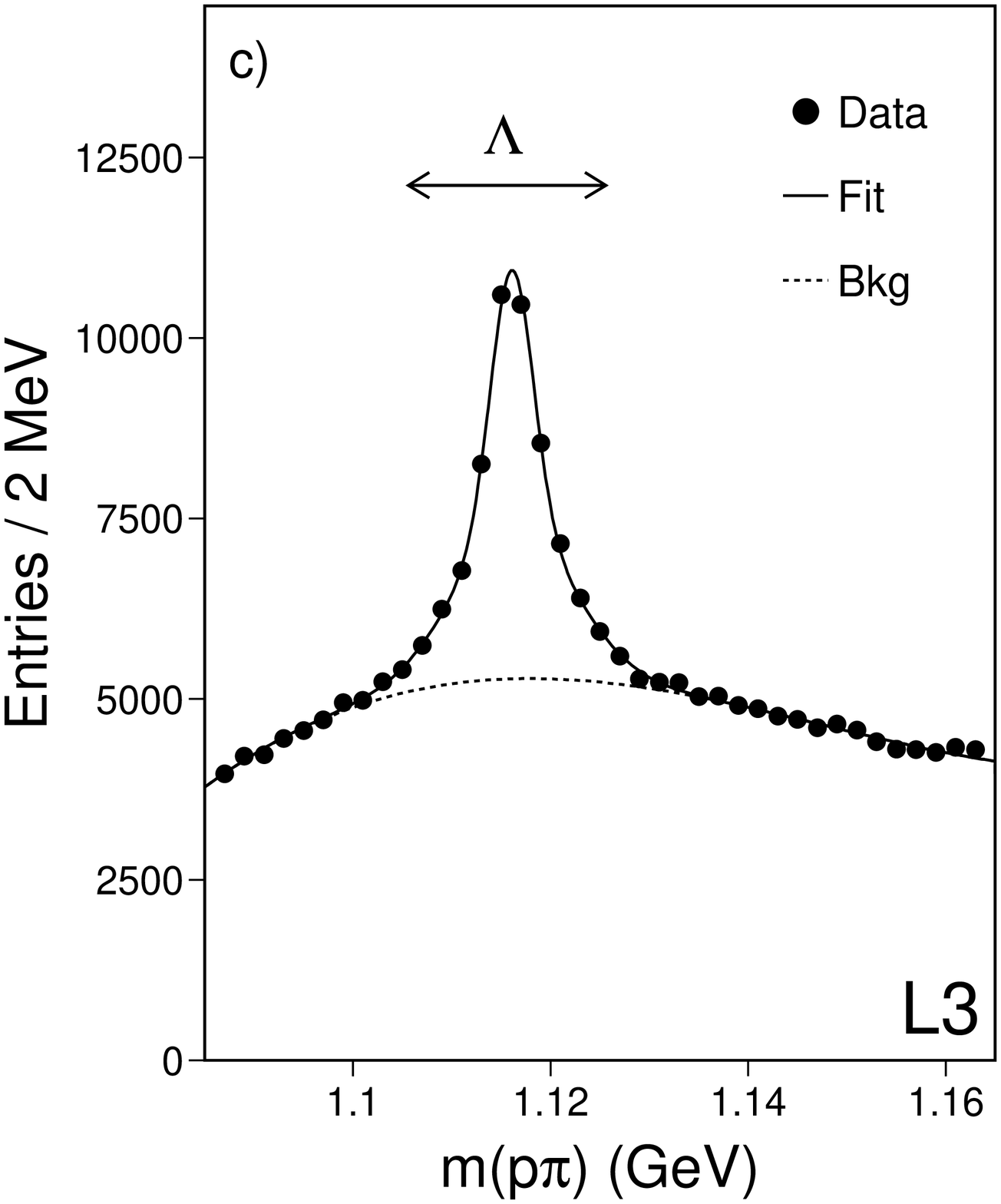,width=7.5cm}
\end{center}
\caption{Distribution of the variables used to select secondary
vertices from $\Lambda$ produced in $\Xi^-$ and $\Omega^-$ decays: 
a) the distance in the transverse plane between the secondary vertex and the 
$\epem$ interaction point, $\rm d_\Lambda$, and b) the angle between the sum $p_t$ vector of 
the two tracks and the direction in the transverse plane between the interaction 
point and the secondary vertex, $\alpha_\Lambda$. All other secondary vertex selection 
criteria are applied. The predictions of the PHOJET Monte Carlo are shown as the full 
line and the contribution due to $\Lambda$ baryons as the 
dashed line. c) The mass of the $\rm p \pi$ system for $\Lambda$ baryons 
produced in $\Xi^-$ and $\Omega^-$ decays. The signal is modelled with two Gaussians and the 
background by a fourth-order Chebyshev polynomial. Only the combinations in the 
$\pm 10 \MeV$ mass window indicated by the arrows are retained to reconstruct 
$\Xi^-$ and $\Omega^-$ baryons.
The Monte Carlo distributions are normalized to the data luminosity.}
\label{lamxicutvtx}
\end{sidewaysfigure}

The $\Xi^-$ baryons are reconstructed by combining $\Lambda$ candidates with 
pion candidates. For the identification of $\Xi^-$ and $\Omega^-$ baryons, different criteria are used 
to select secondary vertices produced by $\Lambda$ decays. The distance $d_\Lambda$ is 
required to be greater than 5 mm and the angle $\alpha_\Lambda$ less than 200 mrad, as 
shown in Figure \ref{lamxicutvtx}a and \ref{lamxicutvtx}b. The other cuts are left unchanged. 
The distribution of the resulting invariant mass $m({\rm p} \pi)$ is displayed in 
Figure \ref{lamxicutvtx}c and shows a 
clear $\Lambda$ peak. The 76000 $\rm p \pi$ combinations that lie in the mass interval 
$ 1.105\GeV < m(\rm p \pi) < 1.125\GeV$ are retained. To reduce the combinatorial background, 
the following criteria are applied: 

\begin{itemize}

\item the distance of closest approach (DCA) in the transverse plane of the $\Xi^-$ decay pion to the 
      $\epem$ interaction point, $d_\pi$, must be greater than 1 mm.

\item the distance in the transverse plane  between the $\Lambda \pi$ vertex and the 
      $\epem$ interaction point, $d_\Xi$, is required to be greater than $d_\Lambda$. 

 \item the angle $\alpha_\Xi$ between the $p_t$ vector of the $\Lambda \pi$ 
       combination and the direction in the transverse plane between the $\epem$ interaction point 
       and the $\Lambda \pi$ vertex has to be less than 100 mrad. 

\end{itemize}

\begin{figure}
\begin{center}
\epsfig{file=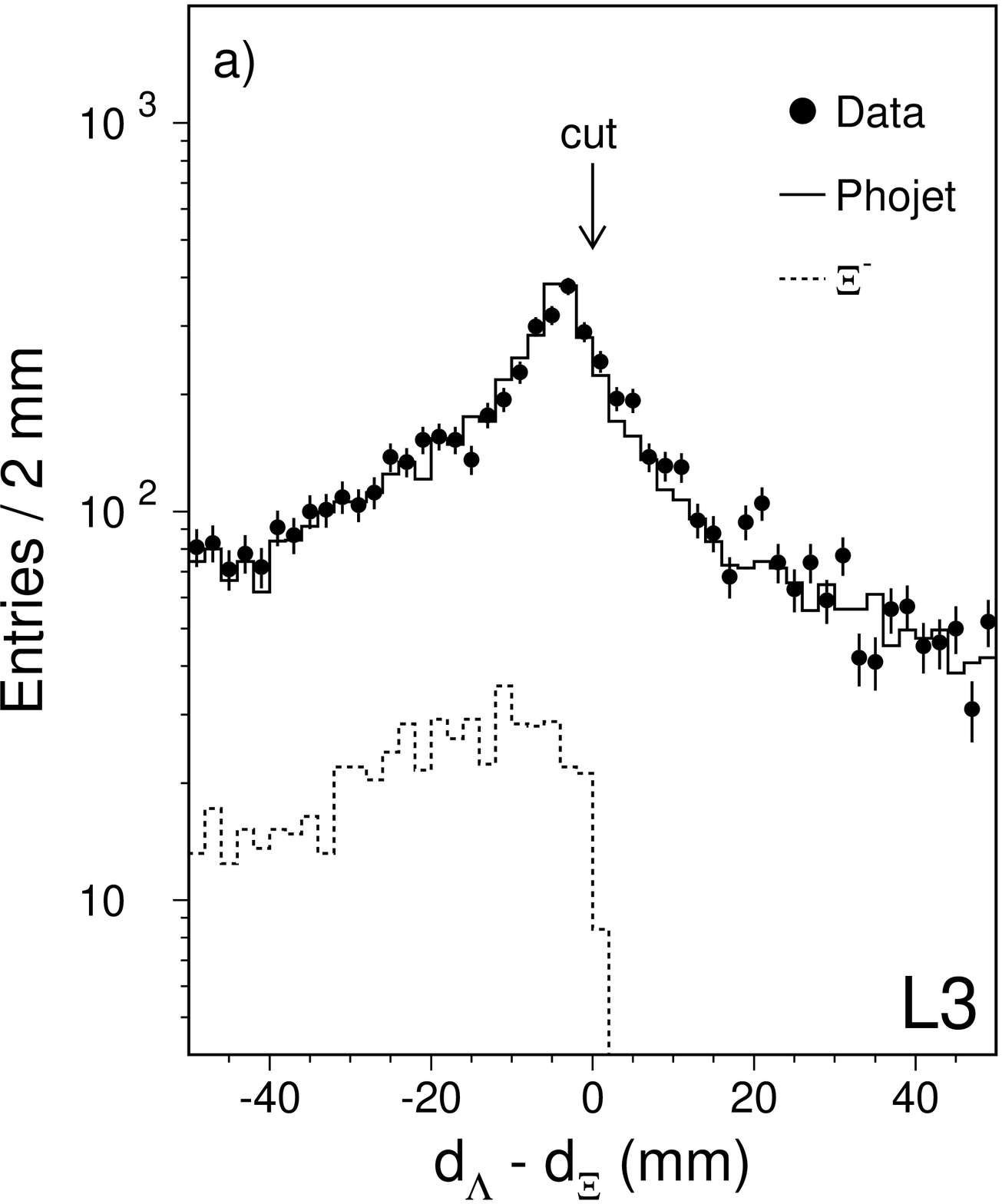,width=7.5cm} 
\epsfig{file=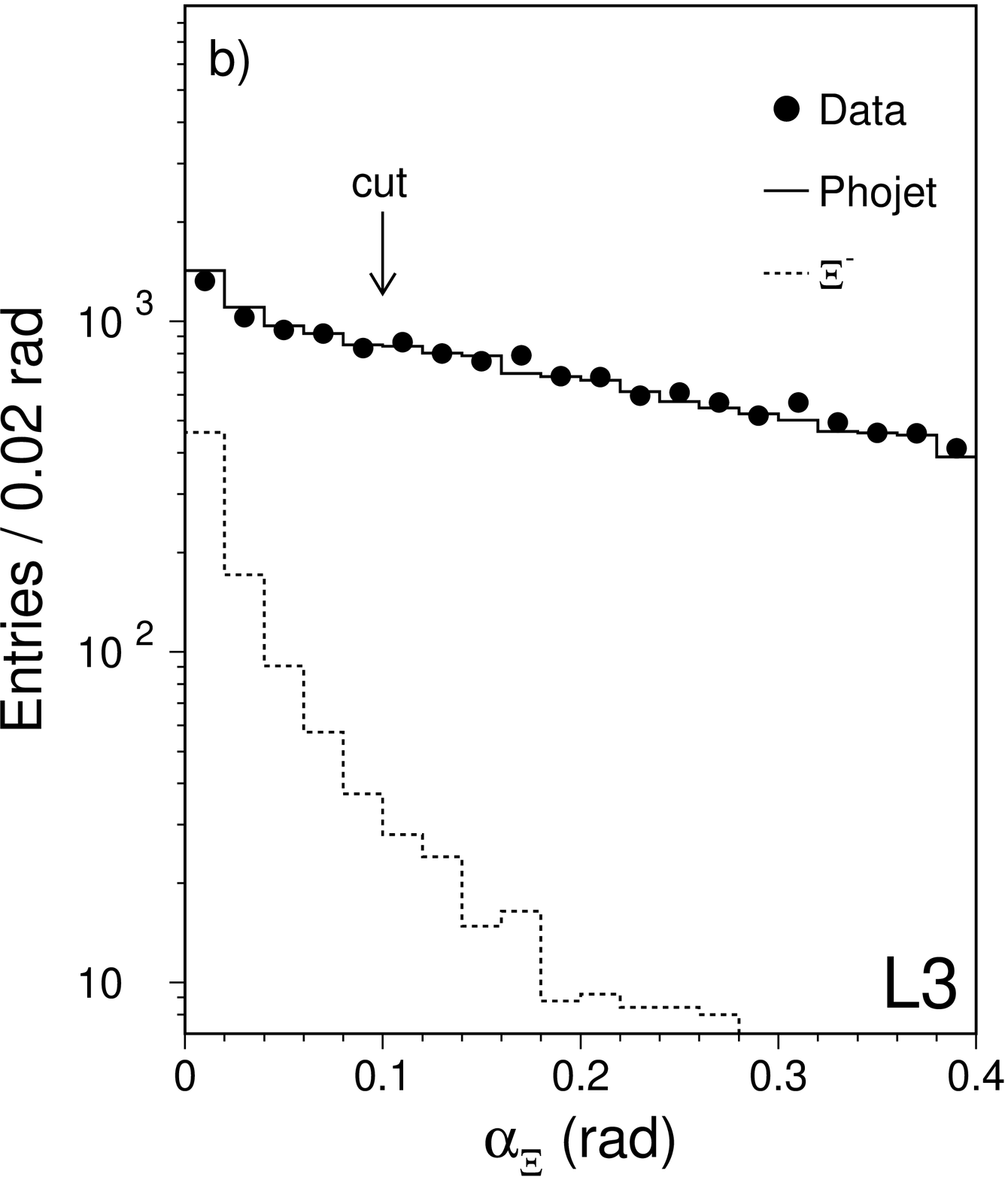,width=7.5cm} 
\end{center}
\caption{Distribution of the variables used for the selection of $\Xi^-$ baryons: a) 
the difference $d_\Lambda -d_\Xi$ of the $\Lambda$ and $\Xi^-$ vertex distances 
from the interaction point and b) the angle $\alpha_\Xi$ between the $p_t$ vector 
of the $\Lambda \pi$ combination and the direction in the transverse plane between the primary interaction point and the $\Lambda \pi$ vertex. 
In each plot, all other selection criteria are applied. The predictions of the PHOJET 
Monte Carlo are shown as the full line and the contribution due to $\Xi^-$ baryons 
as the dashed line. The Monte Carlo distributions are normalized to the data luminosity.}
\label{xicutvtx}
\end{figure}

Distributions of the difference $d_\Lambda-d_\Xi$ and of the angle $\alpha_\Xi$ are 
displayed in Figure \ref{xicutvtx} and exhibit a good agreement with the Monte Carlo 
predictions. 

\begin{figure}
\begin{center}
\epsfig{file=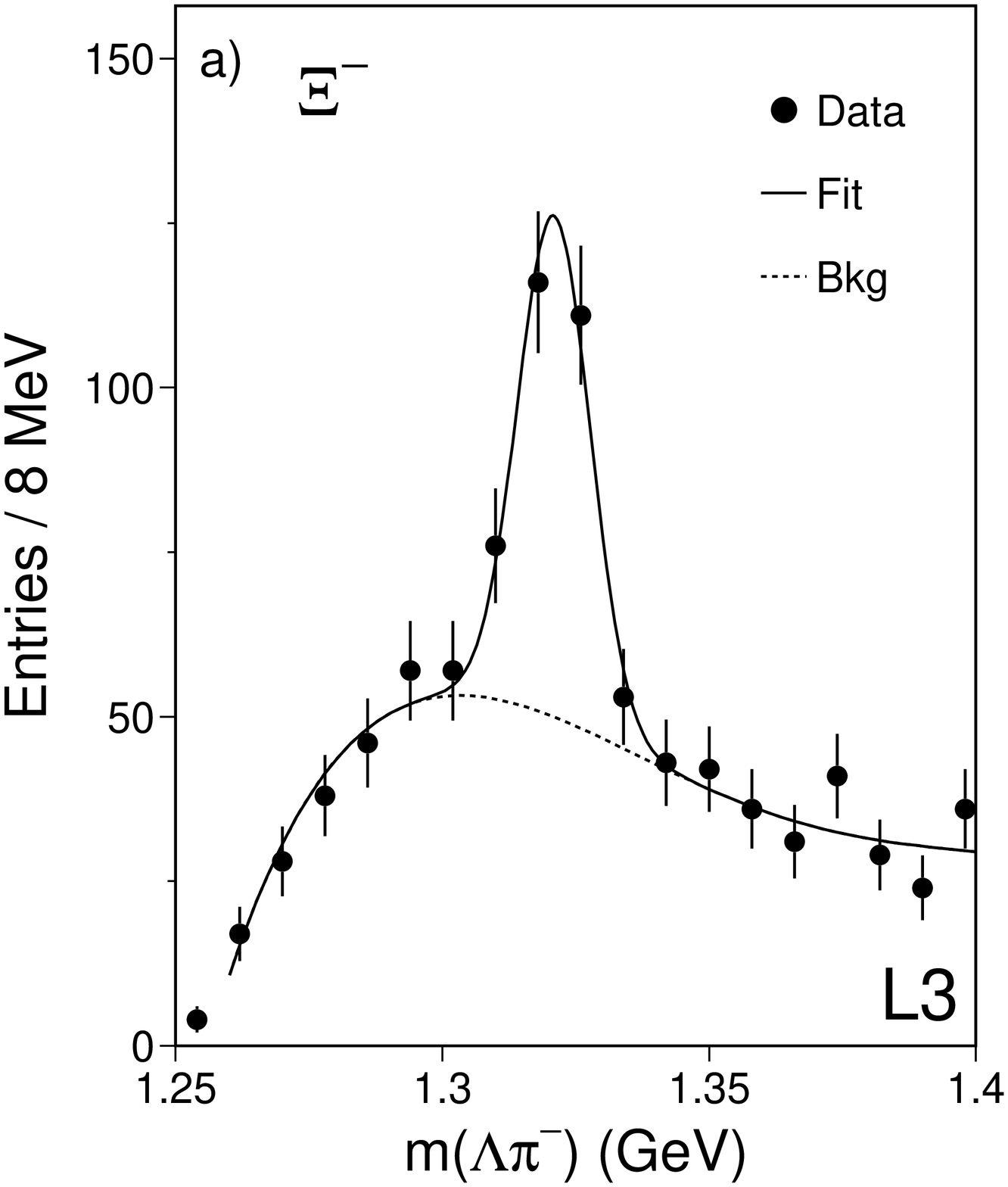,width=7.5cm}
\epsfig{file=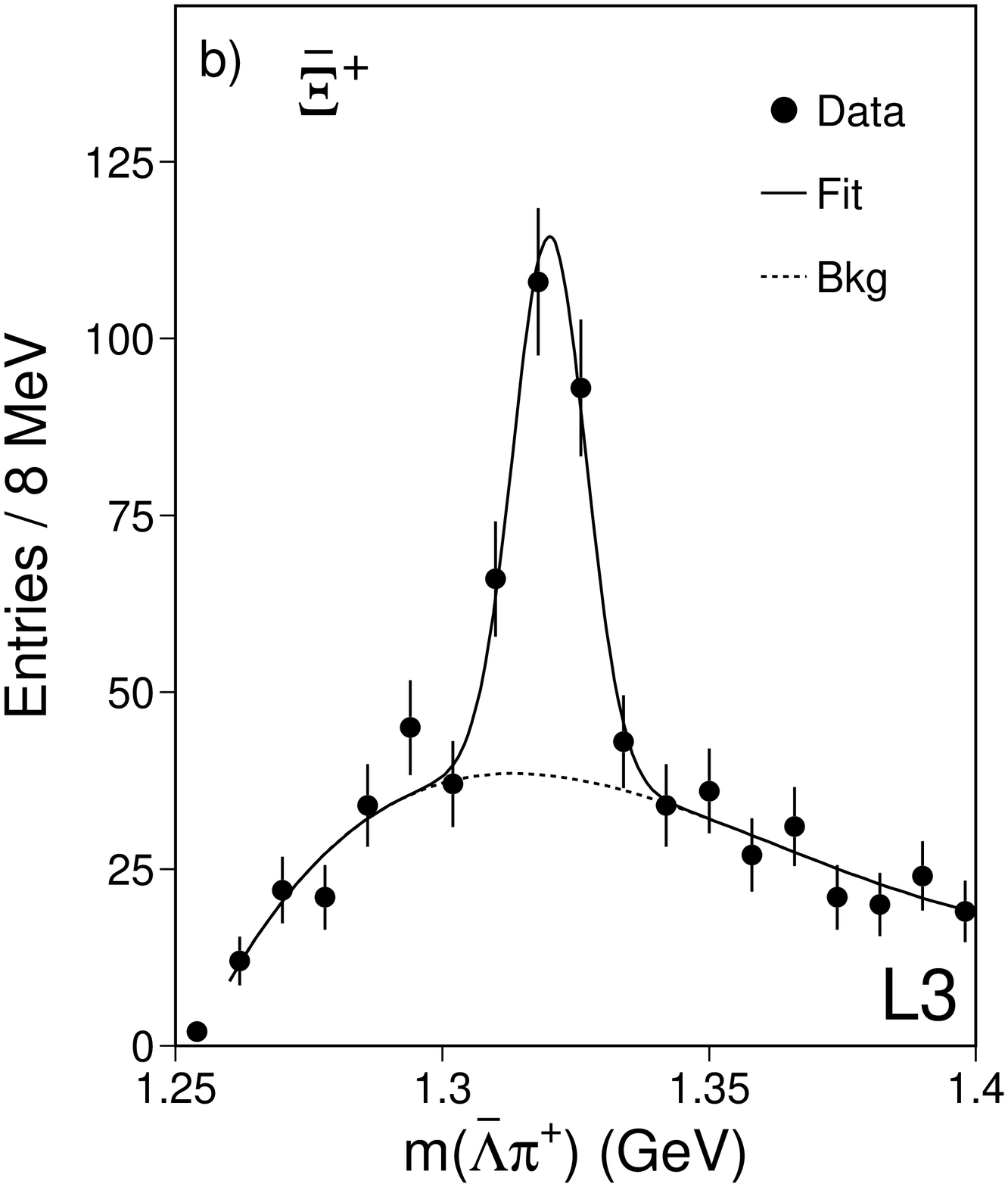,width=7.5cm}
\end{center}
\caption {The mass spectrum of the $\rm \Lambda \pi$ system for a) $\Xi^-$ and b) 
$\overline{\Xi}^+$ candidates for $0.4 \GeV < p_t <2.5\GeV$ and $|\eta|<1.2$. The signal is modelled 
with a Gaussian and the background by a fourth order Chebyshev polynomial.}
\label{xichargedplot}
\end{figure}

\begin{figure}
\begin{center}
\epsfig{file=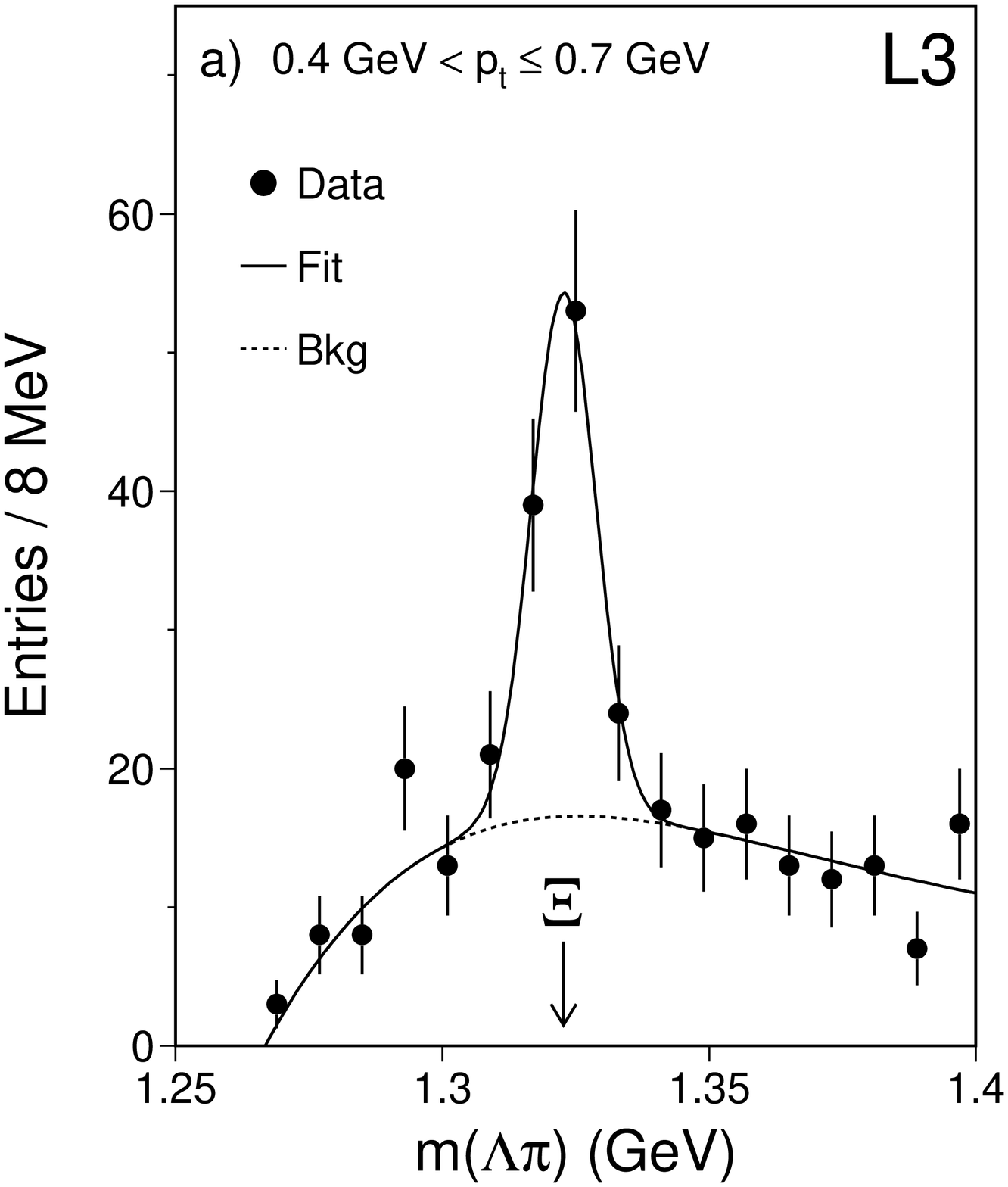,width=7.5cm} \hspace{0.5cm}
\epsfig{file=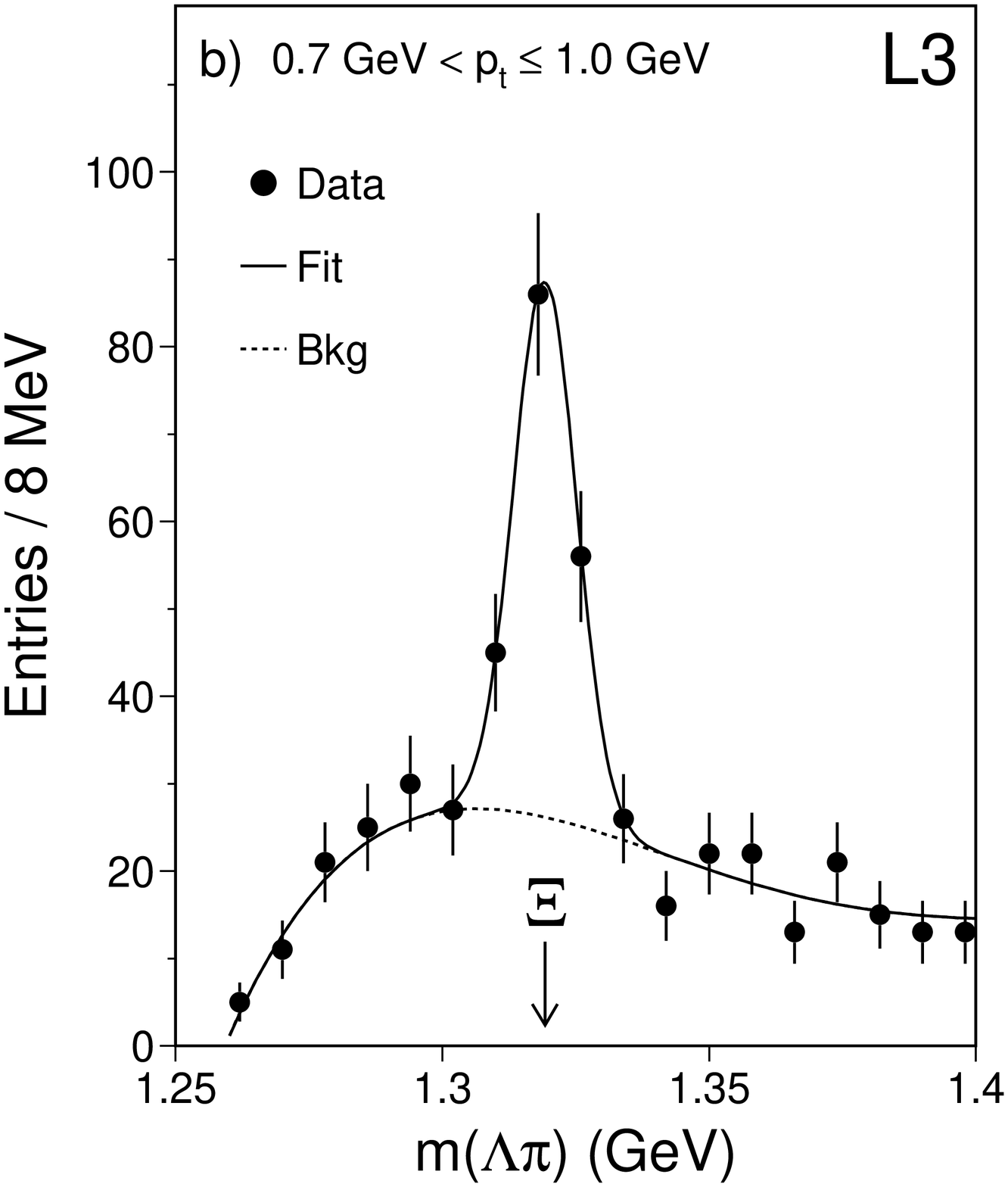,width=7.5cm} \vspace{0.5cm}\\
\epsfig{file=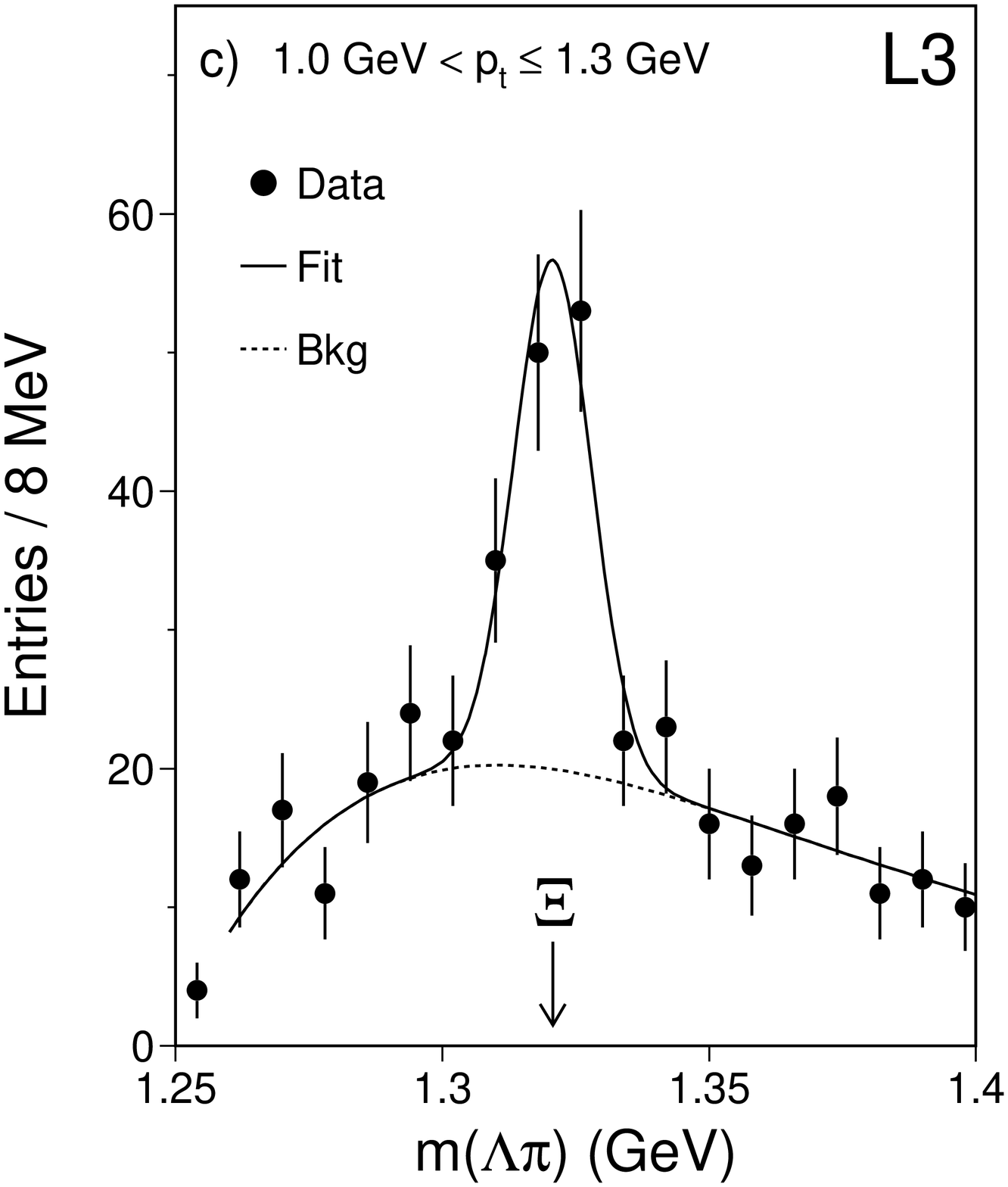,width=7.5cm} \hspace{0.5cm}
\epsfig{file=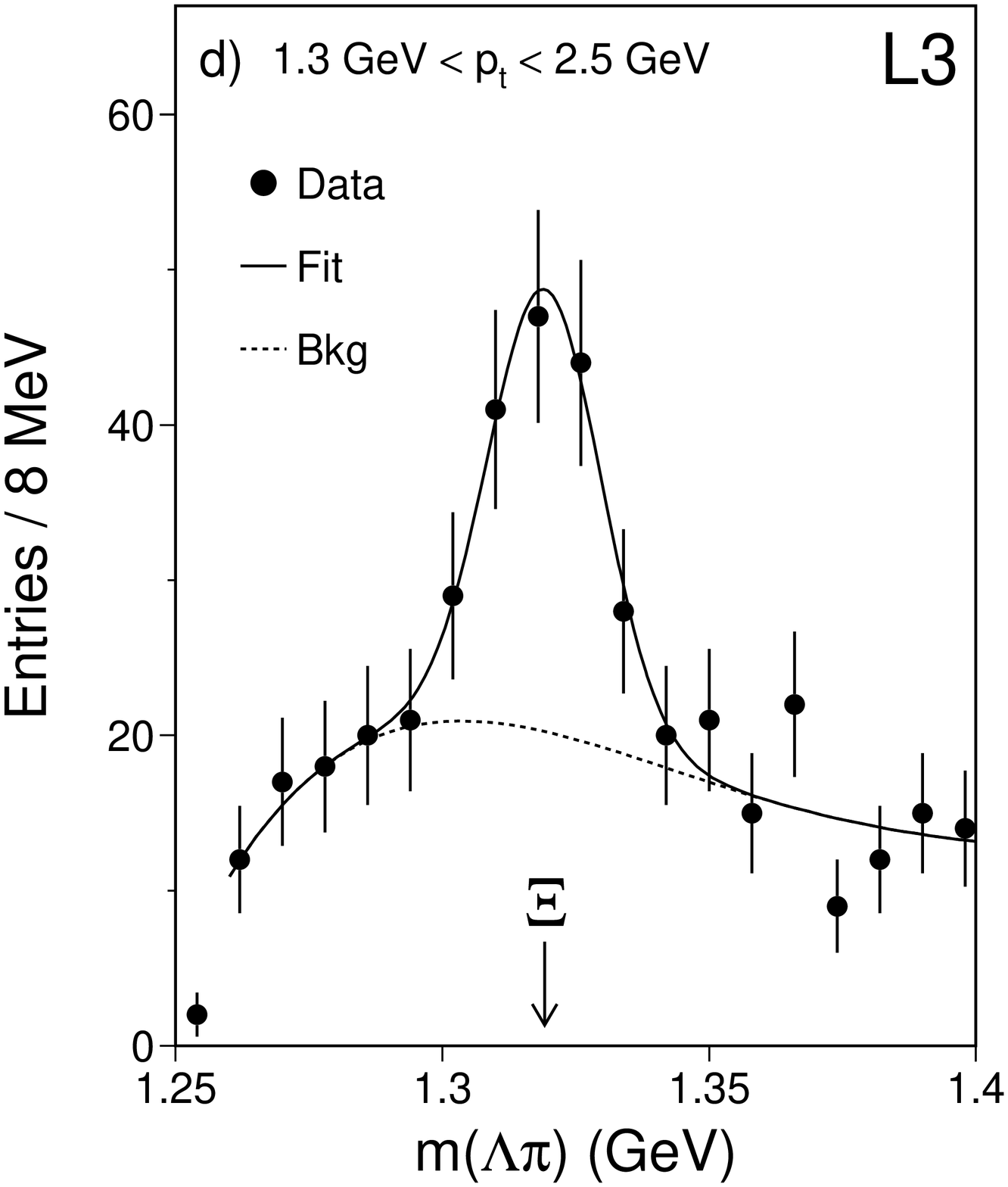,width=7.5cm} 
\end{center}
\caption {The mass of the $\rm \Lambda \pi$ system for a) $0.4 \GeV <
p_t \leq 0.7 \GeV$, b) 
$0.7 \GeV  < p_t \leq 1.0 \GeV $, c) $ 1.0 \GeV  < p_t < 1.3 \GeV $ and d) $1.3 \GeV  < p_t < 2.5 \GeV $. 
The signal is modelled with a Gaussian and the background by a fourth-order Chebyshev polynomial.}
\label{xiplot}
\end{figure}

The distribution of the mass of the $\Lambda \pi$ system, 
$m(\Lambda \pi)$, is displayed in Figure \ref{xichargedplot} for the 
$\rm \Lambda \pi^-$ and $\rm \overline{\Lambda} \pi^+$ combinations, 
respectively. Figure \ref{xiplot} shows the $m(\Lambda \pi)$ spectrum 
for the different $p_t$ bins listed in Table \ref{tabcrosspt}. A clear $\Xi^-$ peak 
is visible over a smooth background, compatible with the measured $\Xi^-$ mass, $m_{\Xi^-} 
= 1321.31 \pm 0.13 \MeV$ \cite{pdg}. The resolution of $m(\Lambda \pi)$, about $7 \MeV$, is well reproduced 
by Monte Carlo simulation. The number of $\Xi^-$ baryons in each $p_t$ and $|\eta|$ 
bin is evaluated by means of a fit to the $ m(\Lambda \pi)$ spectrum in the interval $1.26 \GeV 
< m({\rm p} \pi) < 1.4 \GeV$. The signal is modelled with a Gaussian and the background 
by a fourth-order Chebyshev polynomial. The results are listed in Tables \ref{tabcrosspt} 
and \ref{tabcrosseta}.

\subsection{$\Xi^*(1530)$ selection}

\begin{figure}
\begin{center}
\epsfig{file=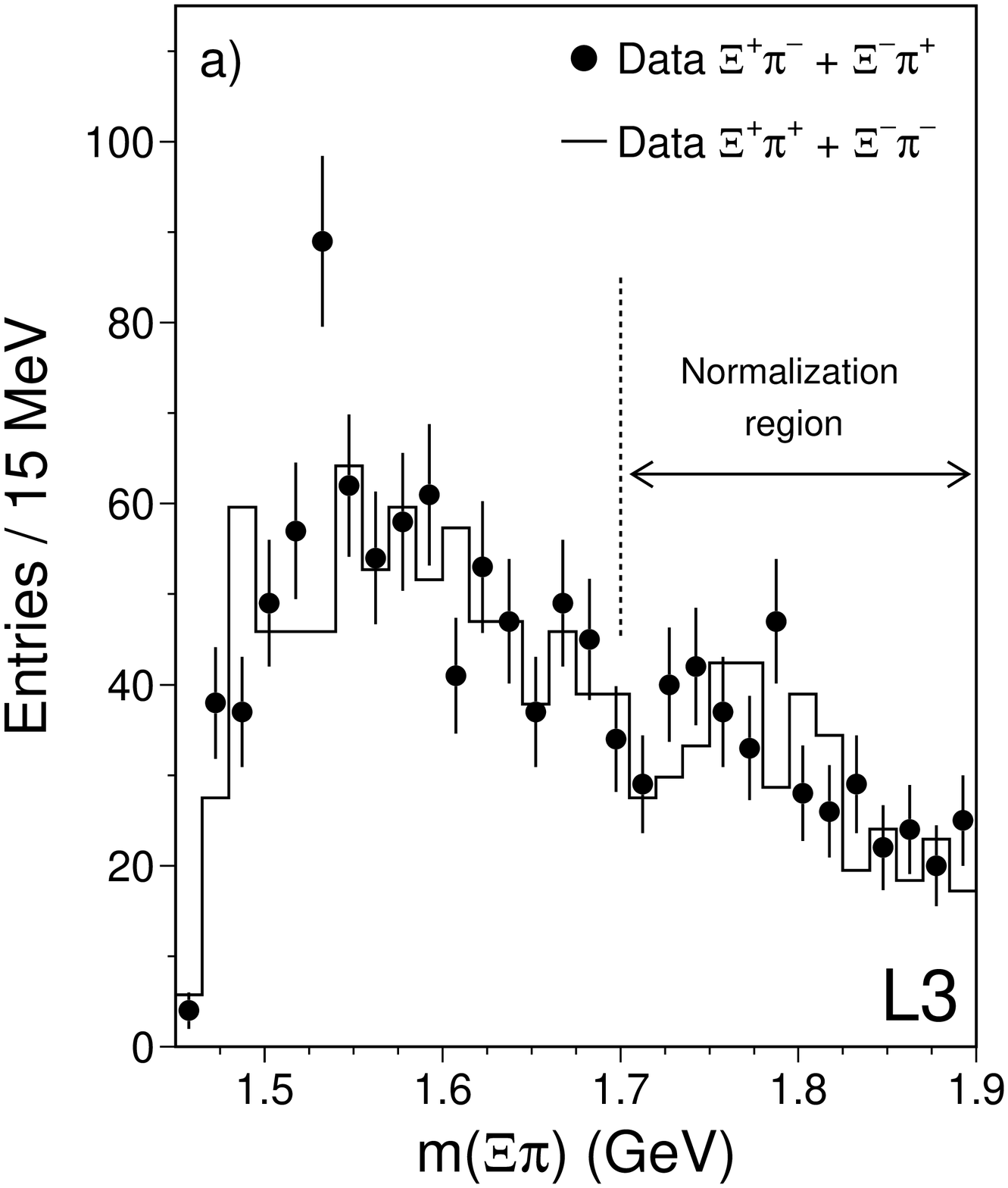,width=7.5cm} \hspace{0.5cm}
\epsfig{file=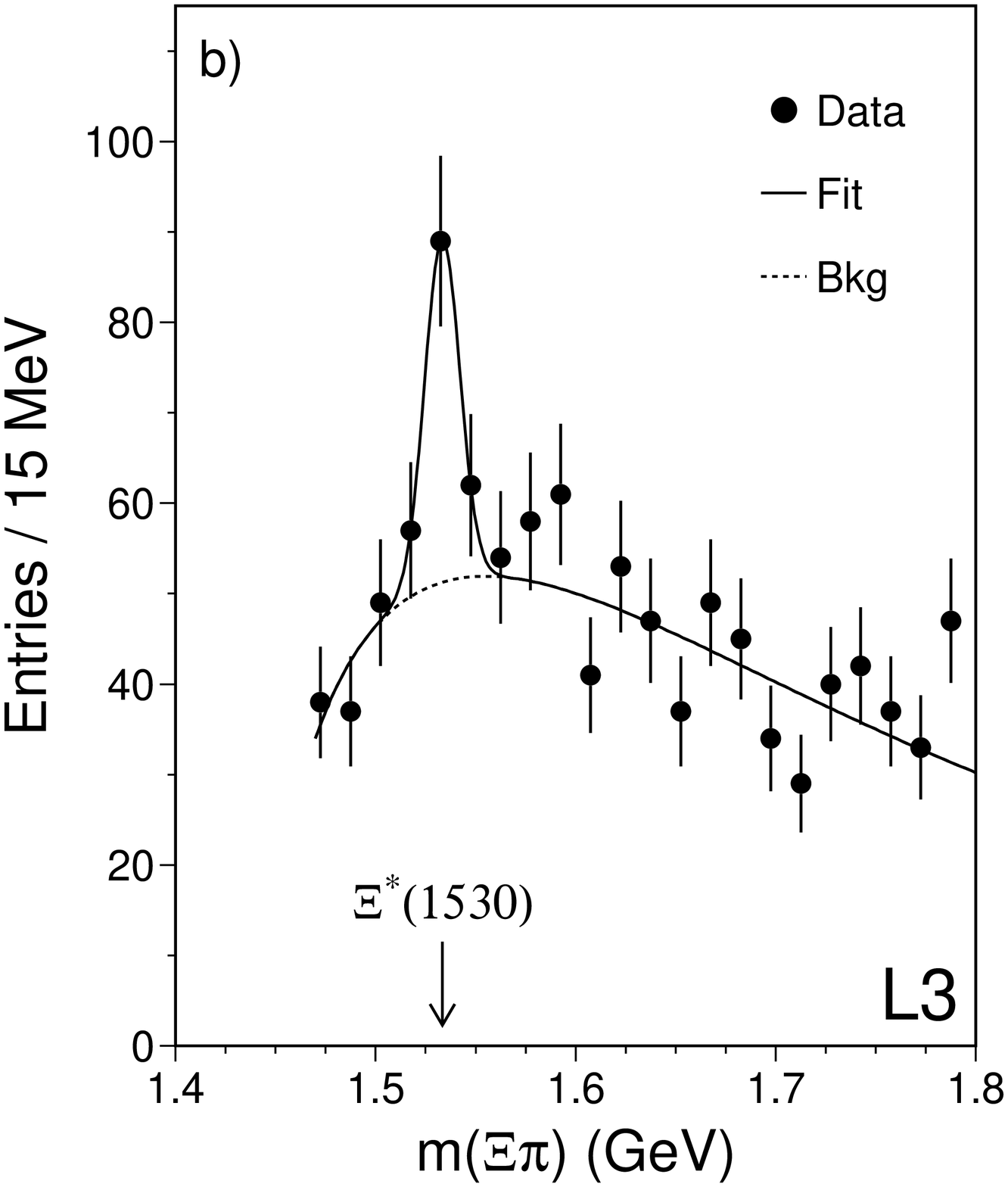,width=7.5cm} 
\end{center}
\caption{a) The mass of the $\Xi \pi$ system for opposite charge ($\Xi^- \pi^+$ and $\overline{\Xi}^+ \pi^-$) 
and same-charge ($\Xi^- \pi^-$ and $\overline{\Xi}^+ \pi^+$) combinations. The number of same-charge 
combinations is normalized to that of opposite-charge combinations in the region $m_{\Xi
\pi} >1.7 \GeV$.  A signal consistent with $\Xi^*(1530)$ production is 
observed. b) Fit to the mass spectrum of the opposite-charge combinations. 
The signal is modelled with a Gaussian and the background by a threshold function.}
\label{xistarplot}
\end{figure}

The $\Xi^*(1530)$ baryons are identified by reconstructing the decays 
$\Xi^*(1530) \ra \Xi^- \pi^+$ and $\Xi^*(1530) \ra \overline{\Xi}^+ \pi^-$. 
The $\Xi^-$ candidates with a mass of $\pm 15 \MeV$ around the nominal 
$\Xi^-$ mass are combined with pion candidates. As the latter are produced 
at the $\epem$ interaction point, their DCA must be less than 5 mm. To compare 
the $\Xi^*(1530)$ production to other strange baryons, the measurement is 
restricted to the kinematical region: $0.4 \GeV < p_t < 2.5\GeV$, $|\eta|<1.2$. The distribution 
of the invariant mass of the $\Xi \pi$ system, $m(\Xi \pi)$, is shown in 
Figure \ref{xistarplot}a for opposite-charge ($\Xi^-\pi^+$ and $\overline{\Xi}^+\pi^-$) and 
same-charge ($\Xi^-\pi^-$ and $\overline{\Xi}^+ \pi^+$) combinations.
The number of same-charge combinations is normalized to that of opposite-charge combinations 
in the region $m(\Xi \pi) > 1.7 \GeV$. An excess corresponding to the $\Xi^*(1530)$ is 
observed in the opposite-charge spectrum close to the nominal mass
$m_{\Xi^*(1530)}=1531.80 \pm 0.32 \MeV$ \cite{pdg}. The number of $\Xi^*(1530)$ baryons 
is determined by means of a fit to the mass spectrum in the region $1.47\GeV < m(\Xi\pi) 
< 1.90 \GeV$, as shown in Figure \ref{xistarplot}b. The signal is modelled with a Gaussian 
and the background is parametrized by a threshold function of the form:
$$a(m_{(\Xi\pi)}-m_0)^b \; \exp\Big\lbrack c(m_{(\Xi\pi)}-m_0) + d(m_{(\Xi\pi)}-m_0)^2 \Big\rbrack$$
where $a,b,c,d$ and $m_0$ are free parameters. The results are given in Table \ref{tabcrosspt}.

\subsection{Search for $\Omega^-$}

\begin{figure}
\begin{center}
\epsfig{file=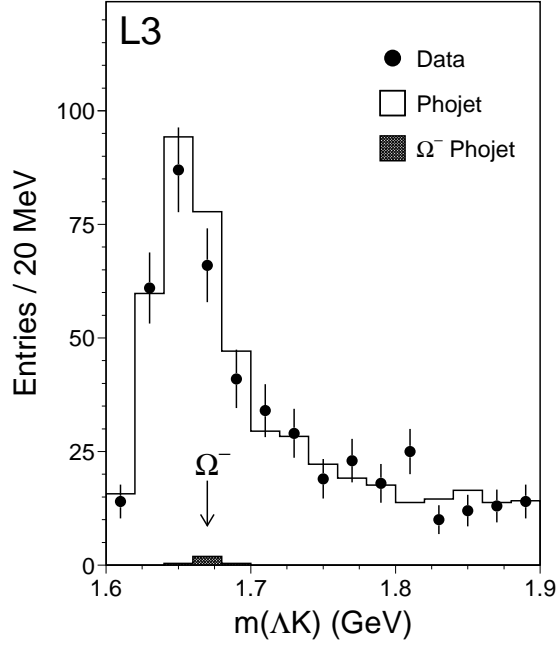,width=7.5cm} 
\end{center}
\caption{The mass of the $\Lambda \rm K$ system. No $\Omega^-$ signal
is observed around the mass of $1.67 \GeV$. The predictions of the PHOJET 
Monte Carlo are shown as the full line and the contribution due to 
$\Omega^-$ baryons as the dashed histogram. The Monte Carlo distributions 
are normalized to the data luminosity.}
\label{omplot}
\end{figure}

Since the topology of the $\Omega^- \ra \Lambda \rm K^-$ decay is similar to that of 
$\Xi^- \ra \Lambda \pi^-$, the selection criteria are identical except that kaon candidates 
are combined with $\Lambda$ candidates instead of pion candidates. The corresponding $m(\Lambda \rm K)$ 
spectrum is displayed in Figure \ref{omplot}. 
No signal is observed around the nominal mass of the $\Omega^-$ baryon, 
$m_{\Omega^-}=1672.45 \pm 0.29 \MeV$ \cite{pdg}. The expected signal, as 
predicted by PHOJET, is found to be almost negligible.

\section{Results and systematic uncertainties}

The cross sections for $\Lambda$, $\Xi^-$ and $\Xi^*(1530)$ production 
are measured for $\wgg > 5 \GeV$, with a mean value $\langle \wgg \rangle =30 \GeV$, and 
a photon virtuality $Q^2 < 8 \GeV^2$ with $\langle Q^2 \rangle \simeq 0.2 \GeV^2$. This kinematical 
region is defined by cuts at Monte Carlo generator level. 

The overall efficiencies for detecting $\Lambda$, $\Xi^-$ and $\Xi^*(1530)$ baryons as a
function of $p_t$ and $|\eta|$ are listed in Tables \ref{tabcrosspt} and
\ref{tabcrosseta}. They include reconstruction and trigger efficiencies as well as 
branching fractions for the decays $\Xi^*(1530) \ra \Xi^- \pi^+$, $\Xi^- \ra \Lambda \pi^-$ 
and $\Lambda \ra {\rm p} \pi^-$: 67\%, 100\% and 64\%, respectively \cite{pdg}. 
The reconstruction efficiencies, which 
include effects of the acceptance and the selection cuts, are calculated with the PHOJET and 
PYTHIA Monte Carlo generators. As both generators reproduce well the shapes of the experimental 
distributions of hadronic two-photon production \cite{l3tot}, the average selection efficiency is 
used. The track-trigger efficiency is calculated for each data-taking period by comparing the 
number of events accepted by the track triggers and the independant calorimetric-energy triggers. The 
efficiency of the higher level triggers is measured using prescaled events. The total trigger 
efficiency varies from 82\% for $p_t<0.4 \GeV$ to 85\% in the high $p_t$ region.

The differential cross sections ${\mathrm d}\sigma/{\mathrm d}p_t$, ${\mathrm d}\sigma/{\mathrm d}p_t^2$ 
and ${\mathrm d}\sigma / {\mathrm d}|\eta|$ for the reactions 
$\epem \ra \epem \Lambda \rm X$ and $\epem \ra \epem \Xi^- \rm X$ are 
given in Table \ref{tabcrosspt} and Table \ref{tabcrosseta}, respectively. 
The mean number of $\Lambda$, $\Xi^-$ and $\Xi^*(1530)$ baryons per hadronic 
two-photon event is measured in the region $0.4 \GeV < p_t< 2.5 \GeV$, 
$|\eta| < 1.2$ and $\wgg > 5\GeV$, as summarized in Table \ref{tabmult}.
Finally, the ratio of $\Lambda$ to $\overline{\Lambda}$ and $\Xi^-$ to $\overline{\Xi}^+$ 
baryons are determined from the respective mass spectra. They are found to be 
$\rm N(\overline{\Lambda}) / N(\Lambda) = 0.99 \pm 0.04$ and 
$\rm N(\overline{\Xi}^+) / N(\Xi^-) = 0.96 \pm 0.14$, where the uncertainties 
are statistical. These results are in  agreement with the value of 1.0 expected for baryon 
pair production in two-photon reactions.

The following sources of systematic uncertainties are considered: selection procedure, 
background subtraction, limited Monte Carlo statistics, Monte Carlo modelling and 
the accuracy of the trigger efficiency measurement. Their contributions to the cross section
measurements are detailed in Table \ref{tabsyst}. 

The uncertainties associated to the selection criteria are evaluated by changing
the corresponding cuts and repeating the fitting procedure. The $\Lambda$ selection 
uncertainty is dominated by the secondary vertex identification while the largest uncertainty 
for the $\Xi^-$ channel, about 9\%, arises from the $\Lambda \pi$ vertex reconstruction. The 
uncertainty due to the $\epem \ra \epem \; hadrons$ event selection is less than 1\%. 

The uncertainty due to background subtraction is assessed by using different background 
parameterizations and fit intervals in the fitting procedure. 
The Monte Carlo modelling uncertainty,  taken as half the relative difference 
difference between PHOJET and PYTHIA, varies between 1\% and 14\% whereas the 
uncertainty associated to the limited Monte Carlo statistics varies from 
1\% for $\Lambda$ to 18\% for $\Xi^*(1530)$. A systematic uncertainty of 2\% is assigned 
to the determination of the trigger efficiency, which takes into account the determination 
procedure and time stability.

\section{Comparison with Monte Carlo and theoretical models}

\begin{figure}
\begin{center}
\epsfig{file=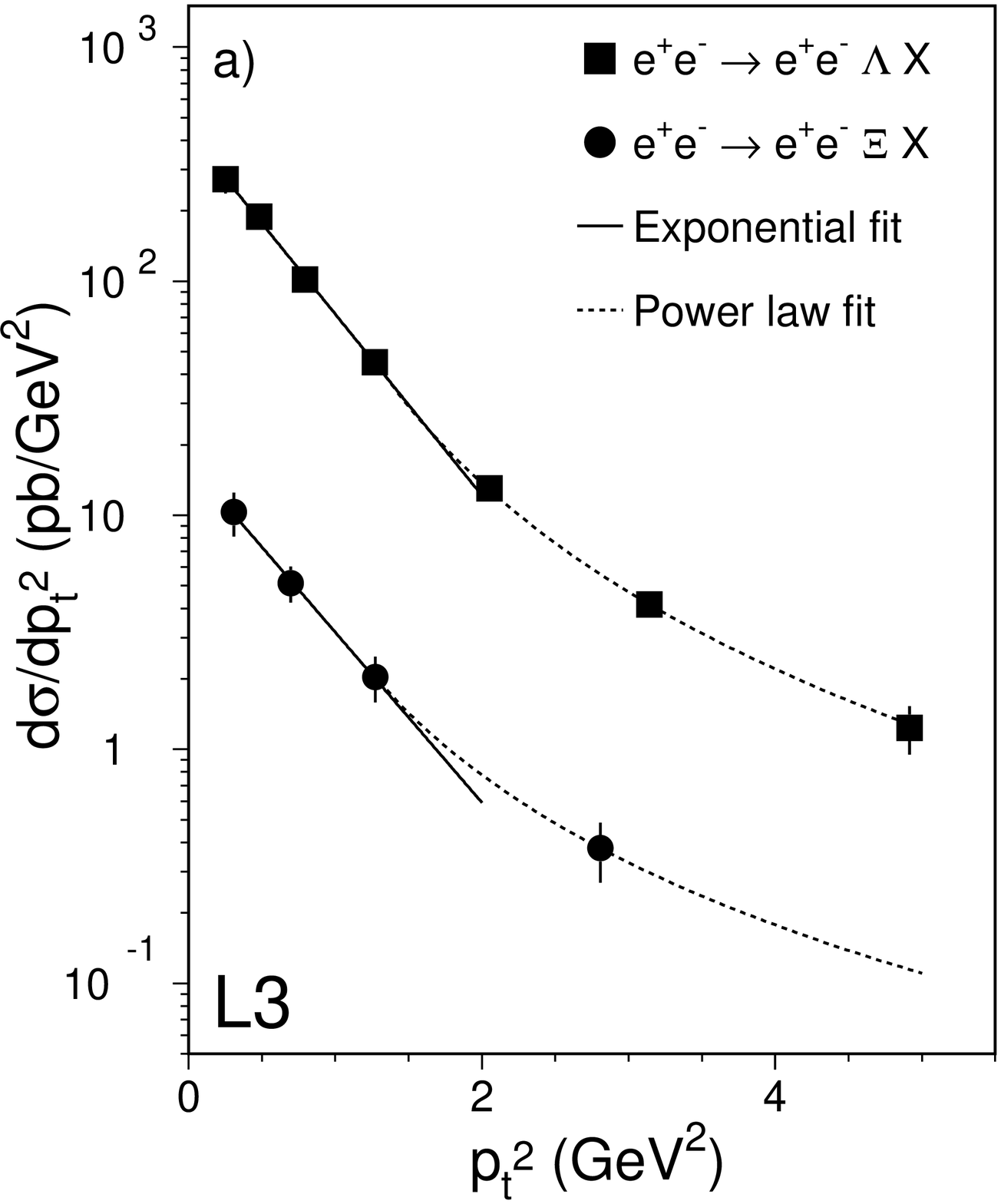,width=7.5cm}
\epsfig{file=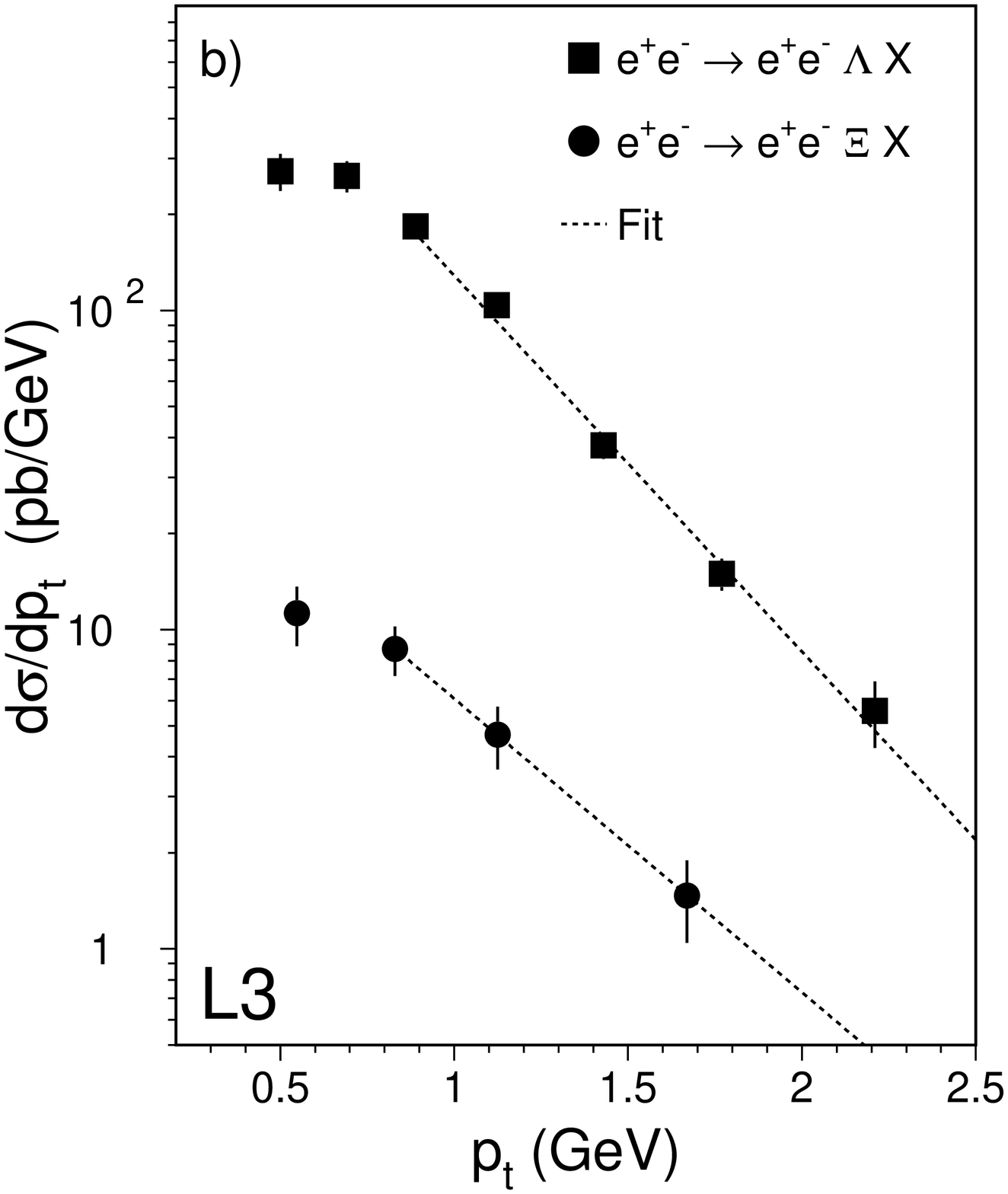,width=7.5cm}
\end{center}
\caption{a) Differential cross sections ${\mathrm d}\sigma / {\mathrm d}p_t^2$ for the $\epem \ra \epem \Lambda \rm
X$ and $\epem \ra \epem \Xi^- \rm X$ processes for $|\eta|<1.2$. b)  Differential cross sections 
${\mathrm d}\sigma / {\mathrm d}p_t$ for the $\epem \ra \epem \Lambda \rm
X$ and $\epem \ra \epem \Xi^- \rm X$ processes for $|\eta|<1.2$. The various fits are described in the text.}
\label{crossfit}
\end{figure}

The differential cross sections ${\mathrm d}\sigma/{\mathrm d}p_t^2(\epem \ra \epem
\Lambda \rm X) $ and  ${\mathrm d}\sigma/{\mathrm d}p_t^2(\epem \ra \epem \Xi^- \rm X) $ 
for $|\eta| < 1.2$ are presented in Figure 
\ref{crossfit}a. The behaviour of the cross sections is 
well described by an exponential of the form $A \exp (-a p_t^2)$ in the region 
$0.16 \GeV^2 < p_t^2 < 1.69\GeV^2$ with $a=1.78\pm 0.16 \GeV^{-2}$ and 
$a=1.68\pm 0.35\GeV^{-2}$ for $\Lambda$ and $\Xi^-$ production, respectively. These two values are 
compatible with each other, as expected from the Lund string model. The region $1.69 \GeV^2 
< p_t^2 < 6.25\GeV^2$ is better reproduced by a power law of the form $A p_t^{-b}$ with 
$b=5.2\pm0.2$ and $b=4.2\pm0.9$ for $\Lambda$ and $\Xi^-$ baryons, respectively.

The differential cross sections ${\mathrm d}\sigma/{\mathrm d}p_t(\epem \ra \epem
\Lambda \rm X) $ and  ${\mathrm d}\sigma/{\mathrm d}p_t(\epem \ra \epem \Xi^- \rm X) $ 
for $|\eta| < 1.2$ are displayed in Figure \ref{crossfit}b. Phase-space 
suppression explains the lower values obtained in the first bins. The behaviour of the cross sections is 
well described by an exponential of the form $A \exp (-p_t/\langle p_t \rangle)$ in the region 
$0.75 \GeV < p_t < 2.5\GeV$ with a mean value $\langle\pt\rangle =
368 \pm 17 \MeV$ for $\Lambda$ production and in the range 
$0.7\GeV < p_t < 2.5\GeV$ with $\langle\pt\rangle =472\pm 98 \MeV$ for $\Xi^-$ formation. 
These values are larger than those obtained for inclusive $\pi^0$ and $\rm K^0_S$ production: 
$\langle\pt\rangle \simeq 230 \MeV$ and $\langle\pt\rangle \simeq 290 \MeV$,
respectively \cite{pz}.

\begin{figure}
\begin{center}
\epsfig{file=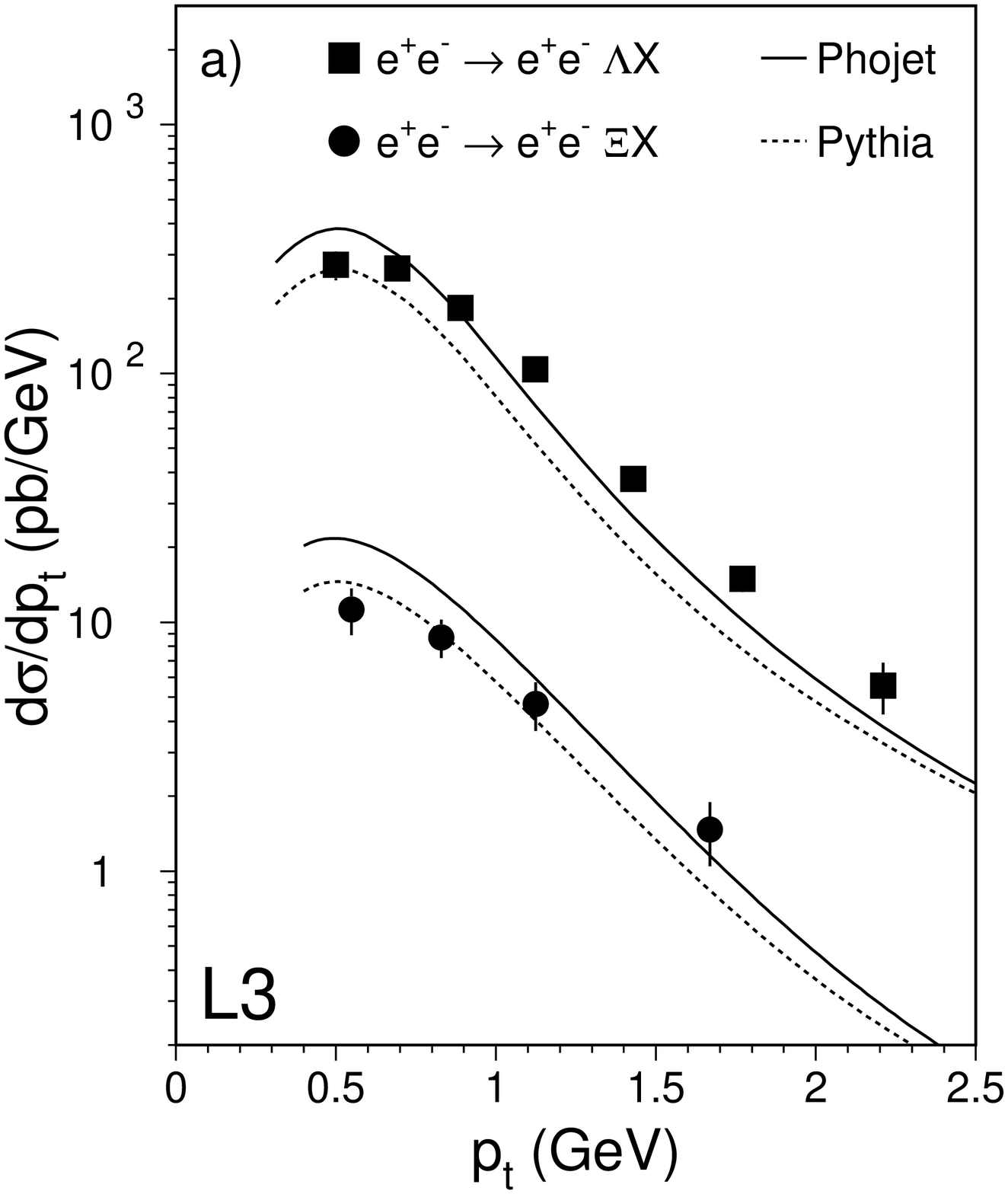,width=7.5cm}
\epsfig{file=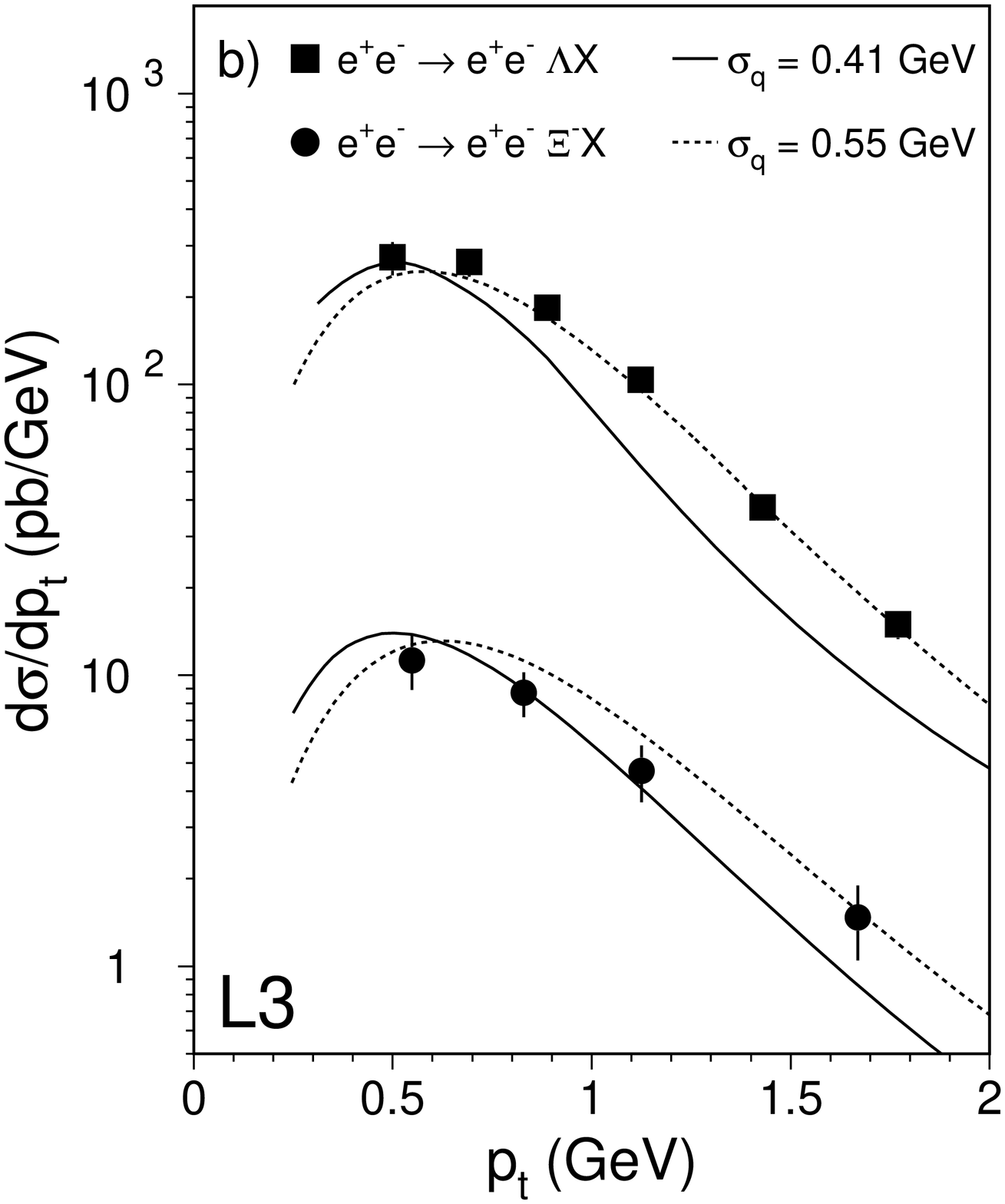,width=7.5cm}
\end{center}
\caption{The differential cross section as a function of $p_t$ for $\Lambda$ and
$\Xi^-$ production for $|\eta|<1.2$ compared with a) the predictions of the PHOJET and 
PYTHIA Monte Carlo, and b) the predictions of the PYTHIA Monte Carlo 
using the default value for the width $\sigma_q$  in the Gaussian $p_t$ distribution 
on primary hadrons $\sigma_q=0.411 \GeV$ as well as the adjusted value 
$\sigma_q=0.55 \GeV$.}
\label{crosscomp}
\end{figure}

The data are compared to the PHOJET and PYTHIA Monte Carlo predictions in 
Figure \ref{crosscomp}a. Both Monte Carlo programs fail to reproduce the shape and 
the normalization of the $\Lambda$ cross section. A better agreement can be achieved by adjusting 
the width $\sigma_q$ in the Gaussian $p_t$ distribution for primary hadrons 
produced during the fragmentation. As shown in Figure \ref{crosscomp}b, the 
predictions obtained with PYTHIA for $\Lambda$ production using a value 
$\sigma_q=0.546 \GeV$ reproduce much better the data than those with the default 
value $\sigma_q=0.411 \GeV$. 
On the other hand, an increase of the width of the $p_t$ distribution of quarks inside 
initial photons does not improve the description of the measured spectra. 
Similar trends are observed with PHOJET. 

\begin{figure}
\begin{center}
\epsfig{file=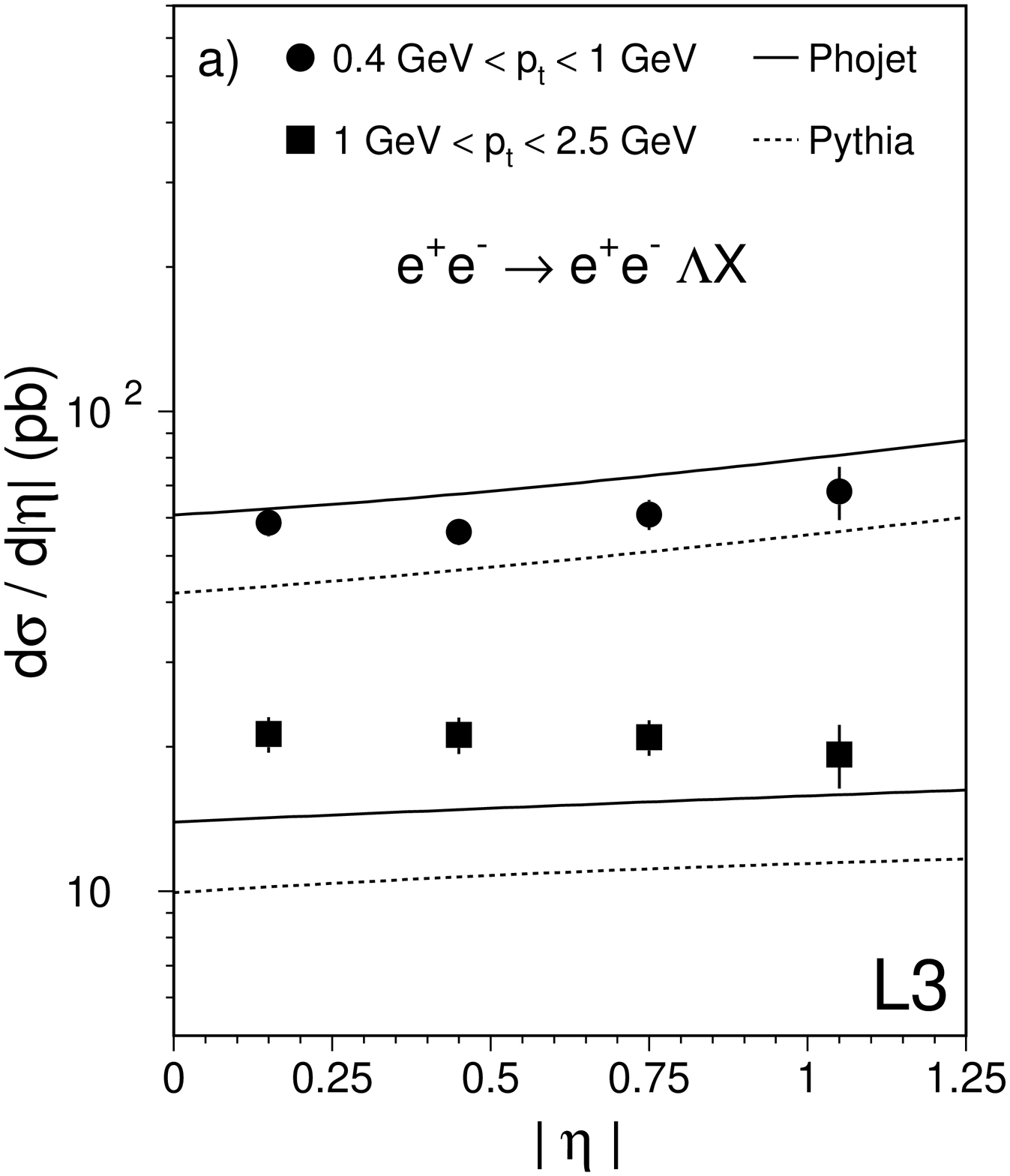,width=7.5cm}
\epsfig{file=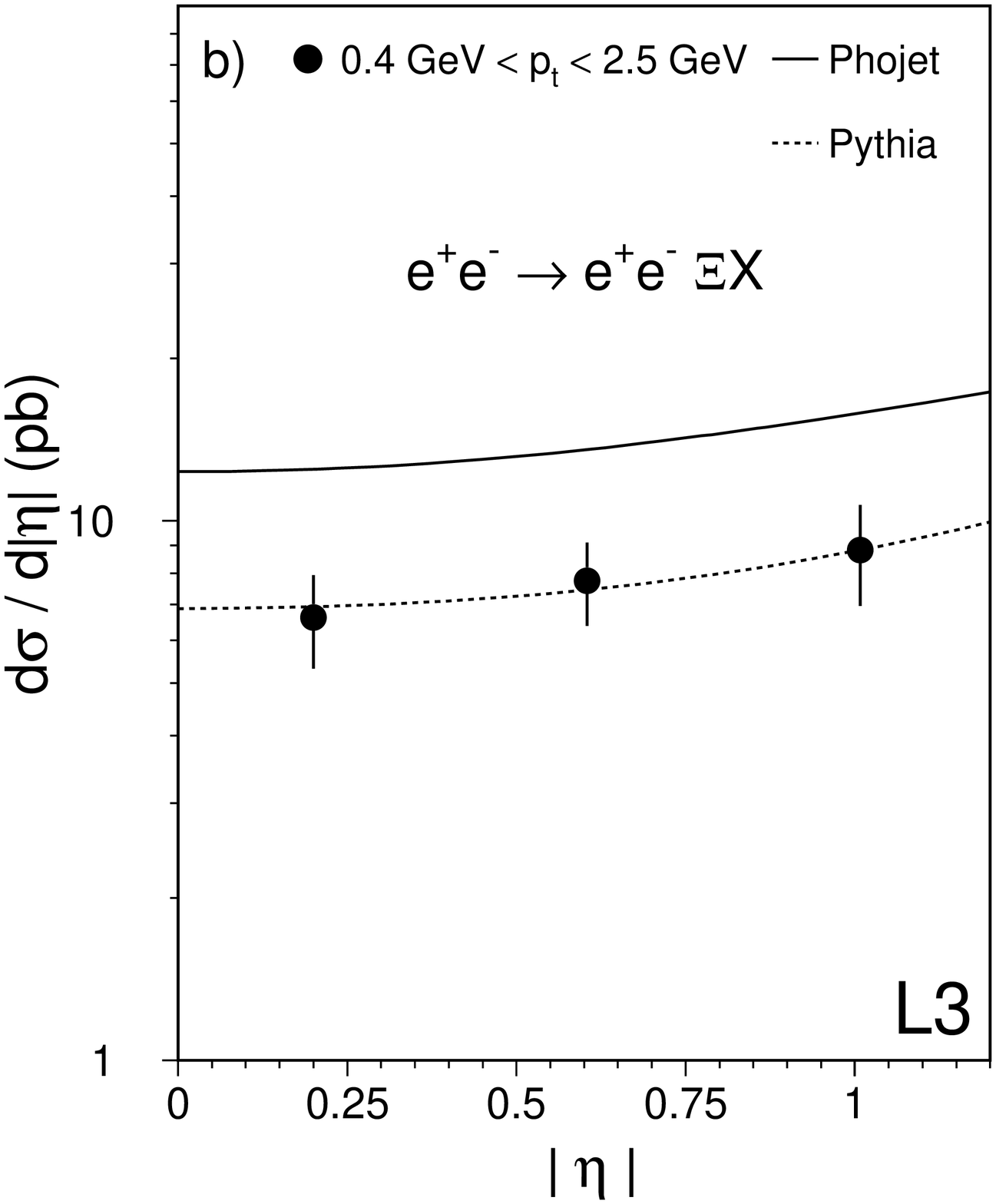,width=7.5cm}
\end{center}
\caption{The differential cross section as a function of $|\eta|$ for a)  inclusive
$\Lambda$ production for $0.4\GeV < p_t < 1.0 \GeV$ and $1.0\GeV < p_t < 2.5 \GeV$ 
and b) for inclusive $\Xi^-$ production for 
$0.4\GeV < p_t < 2.5 \GeV$. The predictions of the PHOJET and PYTHIA programs are 
also shown.}
\label{crosscompeta}
\end{figure}

The differential cross sections as a function of $|\eta|$ are displayed in
Figure \ref{crosscompeta}a and \ref{crosscompeta}b. Both Monte Carlo programs 
describe well the almost-uniform $|\eta|$ shape, while the size of the discrepancy 
on the absolute normalization depends on the $p_t$ range. 

The mean numbers of $\Lambda$, $\Xi^-$ and $\Xi^*(1530)$ baryons per hadronic two-photon 
event are extrapolated from the measured phase-space to the full $p_t$ and $|\eta|$ range 
using the PYTHIA Monte Carlo with the adjusted value $\sigma_q=0.546 \GeV$. 
The uncertainty on the extrapolation factors propagates to the results as an 
additional systematic uncertainty. It arises mainly from the modelling of the two-photon 
reaction and fragmentation processes.
The contribution due to fragmentation, about 7\%, is estimated using an independent 
fragmentation model \cite{indfrag} instead of the Lund string scheme to simulate 
baryon production as well as varying the parameters of these models.
The two-photon modelling uncertainty, taken as the difference between the extrapolation 
factors estimated by PYTHIA and PHOJET, is found to be 3\%. 
The results are summarized in Table \ref{tabmult} together with the predictions 
of PYTHIA and PHOJET. Both Monte Carlos overestimate the $\Lambda$ and $\Xi^-$ 
mean numbers, the discrepancy being less pronounced in PYTHIA than in PHOJET. A better 
agreement is obtained by decreasing the strange-quark suppression factor by a few 
percent. These extrapolated mean numbers can be converted to inclusive cross sections 
by multiplying them for the total two-photon hadronic cross section measured with the same 
detector as discussed in Reference \cite{l3tot}.

\begin{figure}
 \begin{center}
  \epsfig{file=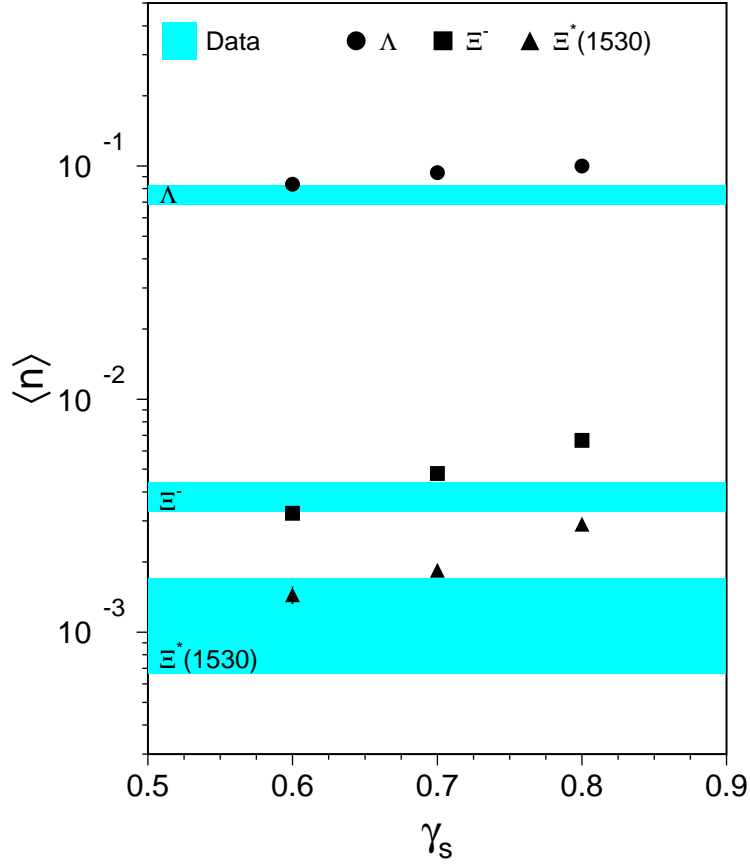,width=10cm}
 \end{center}
\caption{The extrapolated mean multiplicities $\langle n \rangle$ of $\Lambda$, 
$\Xi^-$ and $\Xi^*(1530)$ baryons per two-photon event for $\wgg > 5 \GeV$ (horizontal bars) 
compared to the predictions of the thermodynamical model as a function of the strangeness 
suppression factor, $\gamma_s$, using as energy density the value $\rho=0.4$ (symbols).}
\label{multthermo}
\end{figure}

The predictions of the thermodynamical model of Reference \citen{thermotheo} 
are displayed in Figure \ref{multthermo} as a function of the strangeness 
suppression factor $\gamma_s$, using as energy density the value 
$\rho=0.4$ \cite{thermotheo}. Overall agreement is observed for a $\gamma_s$ 
in the vicinity of $\gamma_s=0.7$, which is similar to the value 
extracted from $\epem$ annihilation events \cite{epemgammas}.

\begin{figure}
 \begin{center}
  \epsfig{file=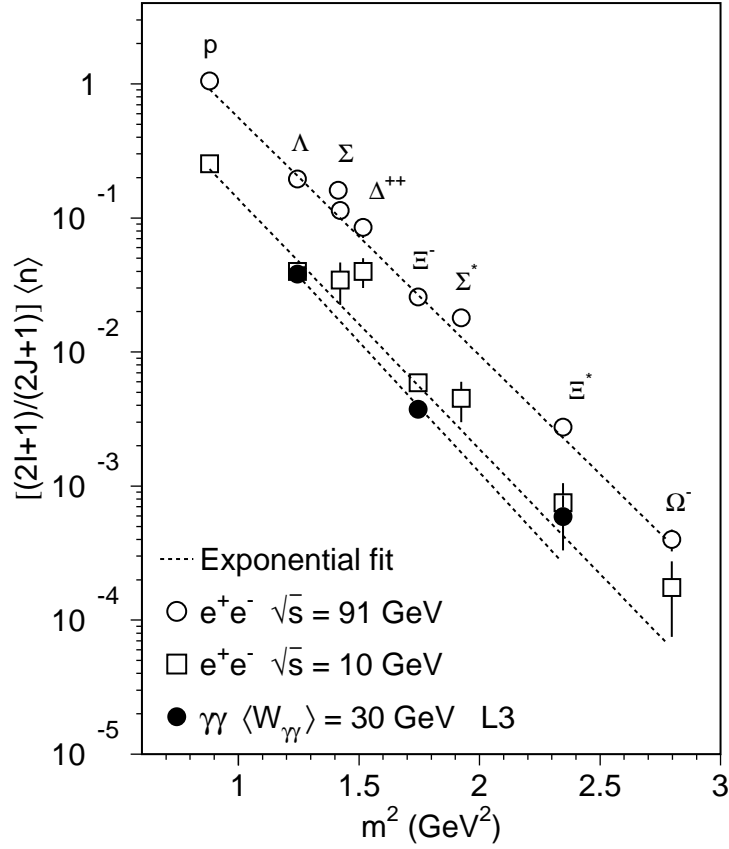,width=10cm}
 \end{center}
\caption{Weighted mean multiplicities $\Big\lbrack (2J+1)/(2I+1) \Big\rbrack \langle n \rangle$ as a
function of the square of the baryon masses measured 
in two-photon reactions (solid circles) together with the data obtained in $\epem$ collisions at 
$\sqrt{s}=10 \GeV$ (open squares) and $\sqrt{s}=91 \GeV$ (open circles). The lines correspond to the fits described in the text.}
\label{mult}
\end{figure}

As suggested in Reference \citen{fitmult}, the mean numbers $\langle n \rangle$ of 
octet and decuplet baryons can also be parametrized by a function of the form:
$$\langle n \rangle =  a \Big\lbrack (2J+1)/(2I+1) \Big\rbrack \exp(-b m^2)$$
where $I,J,m$ are respectively the isospin, total angular momentum and mass of
the baryons and $a,b$ are free parameters. 
The ratios $\Big\lbrack (2J+1)/(2I+1) \Big\rbrack \langle n \rangle $ are displayed in Figure \ref{mult} 
together with the data measured in $\epem$ collisions at $\sqrt{s}=10 \GeV$ and 
$\sqrt{s}=91 \GeV$ \cite{pdg}. A fit to our data gives $b=4.5 \pm 0.9$, a value 
compatible with the exponents $b=4.3 \pm 0.6$ and $b=4.0 \pm 0.1$ obtained 
for $\epem$ reactions at $\sqrt{s}=10\GeV$ and $\sqrt{s}=91 \GeV$, respectively.
This provides evidence for the universality of fragmentation processes in two-photon and
$\epem$ reactions.

\section{Search for the $\theta^+$ pentaquark}

A $\theta^+$\footnote{If not stated otherwise, 
the symbols $\theta^+$ refers to $\theta^+$ and $\overline{\theta}^{-}$ 
as both charge-conjugate final-states are analysed.} Monte 
Carlo consisting of pentaquarks mixed with large samples of 
hadronic two-photon events is used to study the kinematics of the $\theta^+$ decay 
products and estimate the signal efficiency. In this Monte Carlo the pentaquarks are 
generated with a mass of $1.54 \GeV$ and a width of $1 \MeV$. The transverse momentum 
and pseudo-rapidity distributions predicted by the PYTHIA event generator 
for inclusive $\Xi^*(1530)$ production are taken to model the unknown 
distributions of the pentaquarks. As these distributions are very similar for 
different low mass baryons, the simulated acceptance is expected to be 
quite insensitive to the choice of a particular baryon. The $\theta^+ \ra \rm p 
\kos$ decay is assumed to be isotropic. 
The $\kos \rm  p$ and $\kos \rm  \overline{p}$ mass distributions, calculated 
at the generator level, show no evidence of a narrow peak close to $1540 \MeV$. This 
indicates that no reflections from known decay modes could generate a 
fake $\theta^+$ signal.

\subsection{Event selection}

The $\theta^+$ candidates are reconstructed using the decay $\theta^+ \ra \rm p \kos$.
Events are selected if they contain at least one track coming from the $\epem$  interaction point 
and a $\kos$ decaying into $\pi^+ \pi^-$ at a secondary vertex. 
The $\kos$ selection proceeds as follows:

\begin{itemize}

 \item Each track is required to have more than 12 hits out a maximum 
 of 62 and $p_t > 100 \MeV$.

 \item The d$E$/d$x$ measurement of both pion candidates must be 
 consistent with this hypothesis with a probability greater than 0.01.

 \item The distance in the transverse plane between the secondary vertex and 
 the $\epem$ interaction point must be greater than 5 mm. 

 \item The angle between the sum $p_t$ vector of 
 the two tracks and the direction in the transverse plane 
 between the primary interaction point and the secondary vertex must be 
 less than 75 mrad. 

\end{itemize}

\begin{figure}
\begin{center}
\epsfig{file=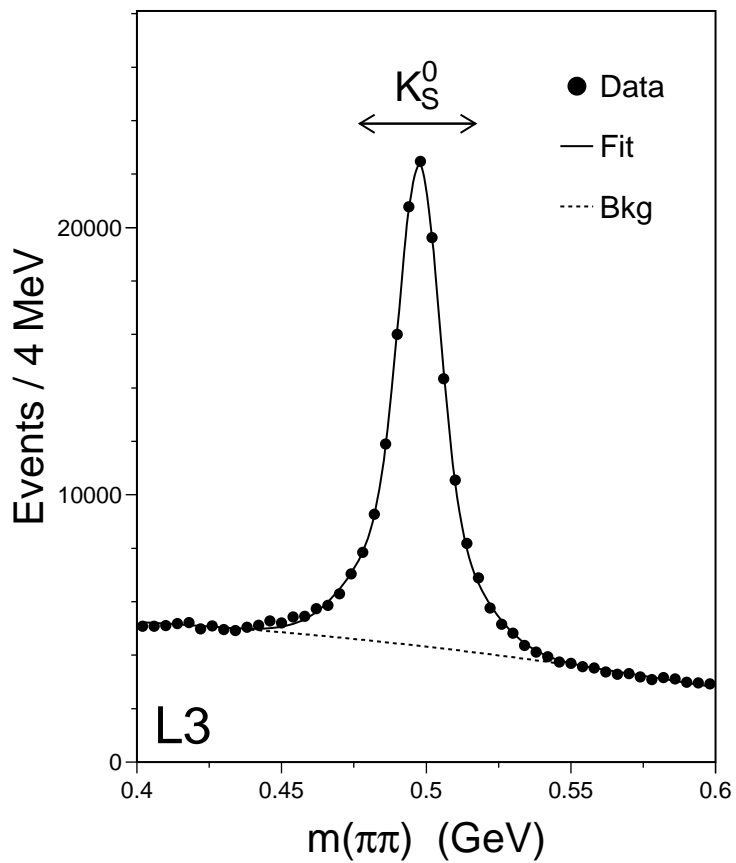,width=10cm}
\end{center}
\caption{The effective mass of the $\pi\pi$ system. The signal is modelled 
with two Gaussian functions and the background by a second-order polynomial. 
About 140000 $\kos$ candidates are found in a $\pm 20 \MeV$ window around the 
central value of the peak and used in the search for the $\theta^+$ pentaquark.}
\label{massk0s}
\end{figure}

The distribution of the effective mass of the $\pi \pi$ system, $m(\pi \pi)$, 
is shown in Figure \ref{massk0s}. A clear $\kos$ peak is present over a smooth
background. The spectrum is fitted with two Gaussian functions and a second order 
polynomial for the background. A central value of $497.6 \pm 0.1 \MeV$ is 
obtained for the peak, which agrees with the expected value of $497.7 \MeV$ \cite{pdg}. 
To search for $\theta^+$, the $\kos$ candidates are selected within an interval of 
$\pm 20 \MeV$ around the central value of the peak, corresponding to 140000 $\kos$ 
candidates with a purity of 69\%. 

\begin{figure}
\begin{center}
\epsfig{file=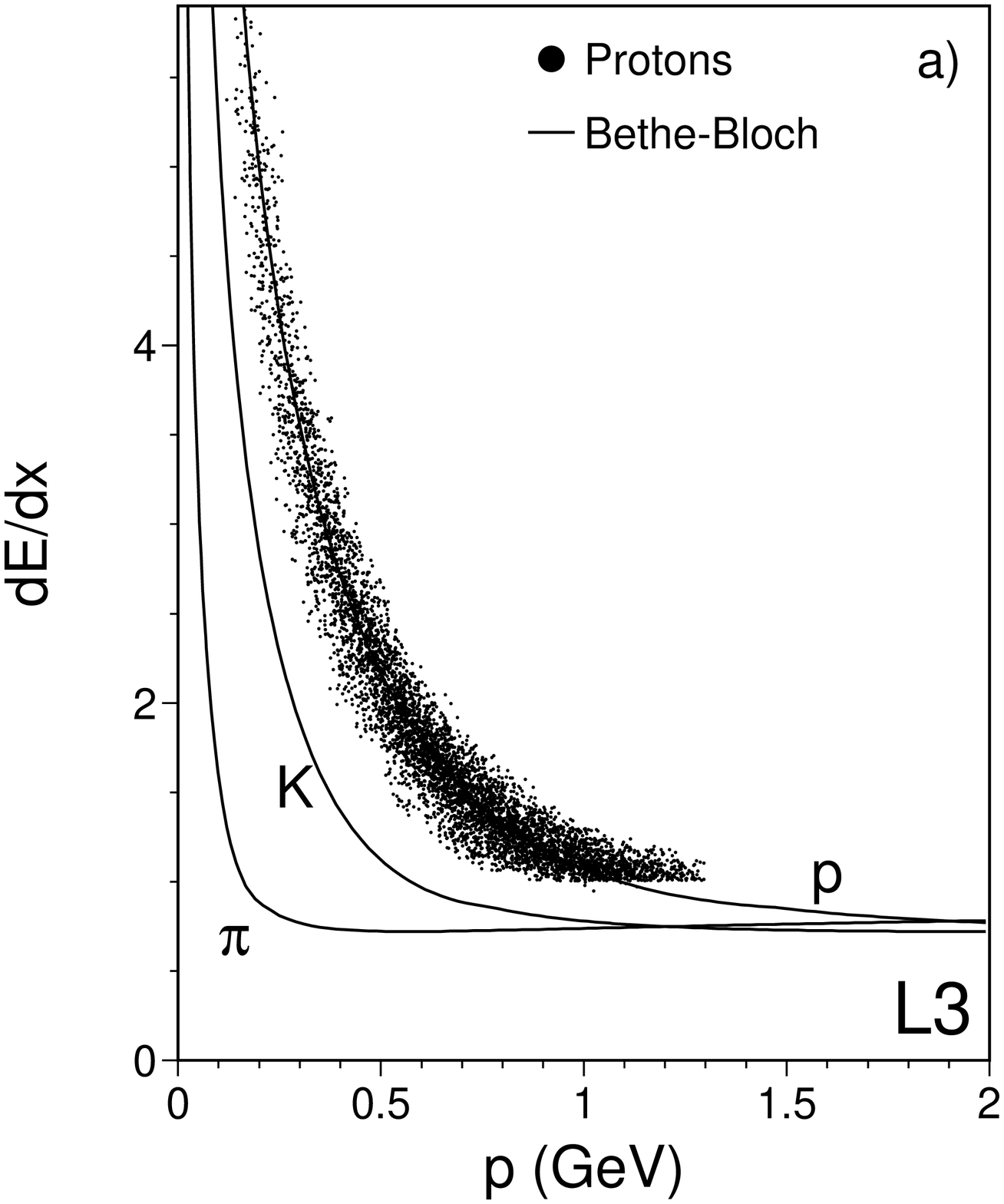,width=7.5cm}
\epsfig{file=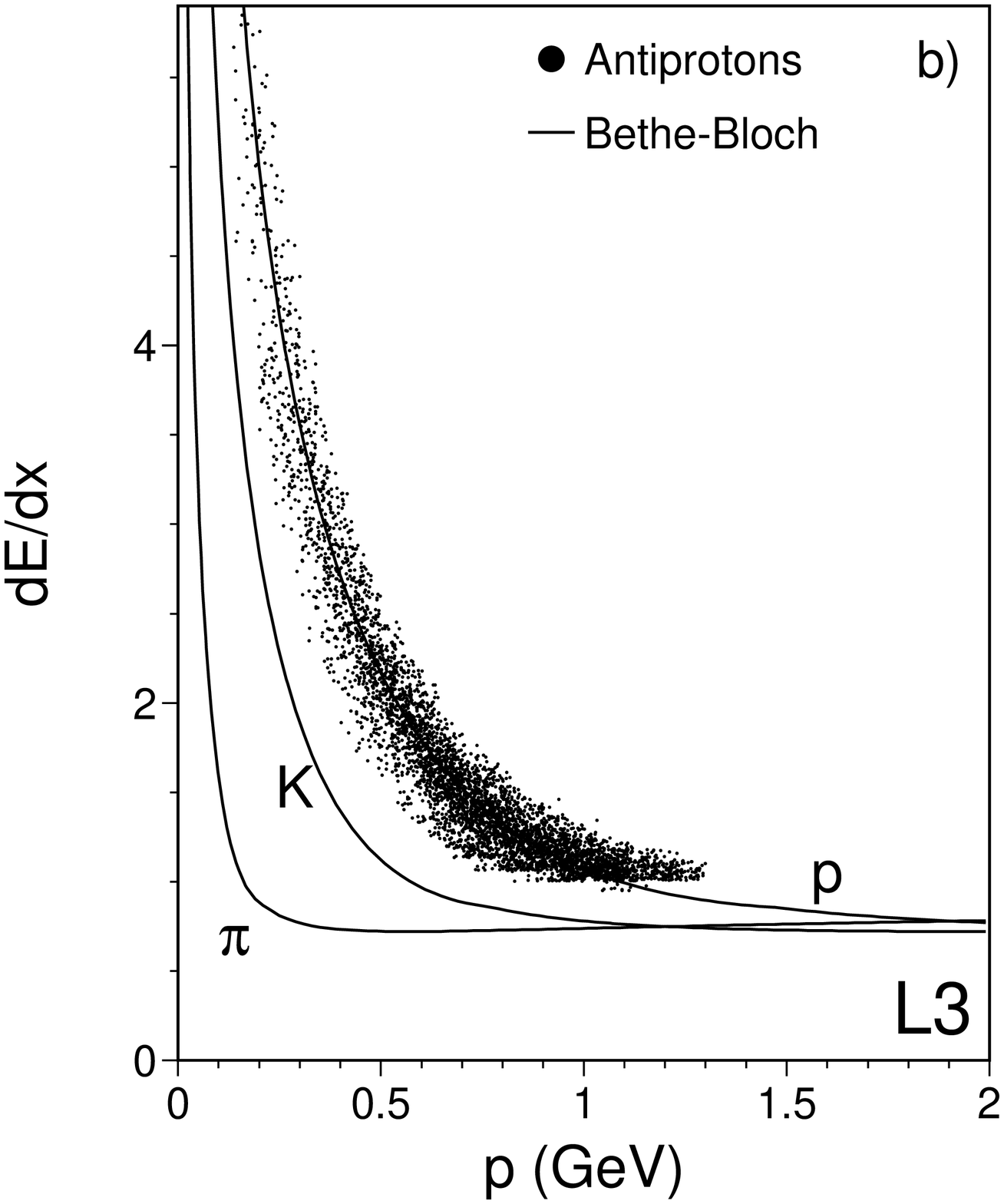,width=7.5cm}
\end{center}
\caption{The d$E$/d$x$ measurement as a function of the momentum for the selected
a) protons and b) antiprotons, together with the calculations of the Bethe-Bloch formula for
protons, kaons and pions.}
\label{protsel}
\end{figure}

The proton and antiproton candidates are tracks with at least 30 hits out of a 
maximum of 62 and a DCA less than 5 mm. In addition, the d$E$/d$x$ measurement in 
the central tracker must be consistent with the proton hypothesis with a probability 
greater than 0.05. To reject kaons and pions, the probability of each of these 
hypothesis must be less than 0.01. A total of 7131 selected protons and 5340 
antiprotons are selected. Their d$E$/d$x$ measurement is shown in Figure \ref{protsel} 
together with theoretical the expectations based on the Bethe-Bloch formula. The 
purity of the selected sample is greater than 96\%. The excess of protons, due to secondary 
interactions between particles and the detector, is well reproduced by the Monte Carlo. 

\begin{figure}
\begin{center}
\epsfig{file=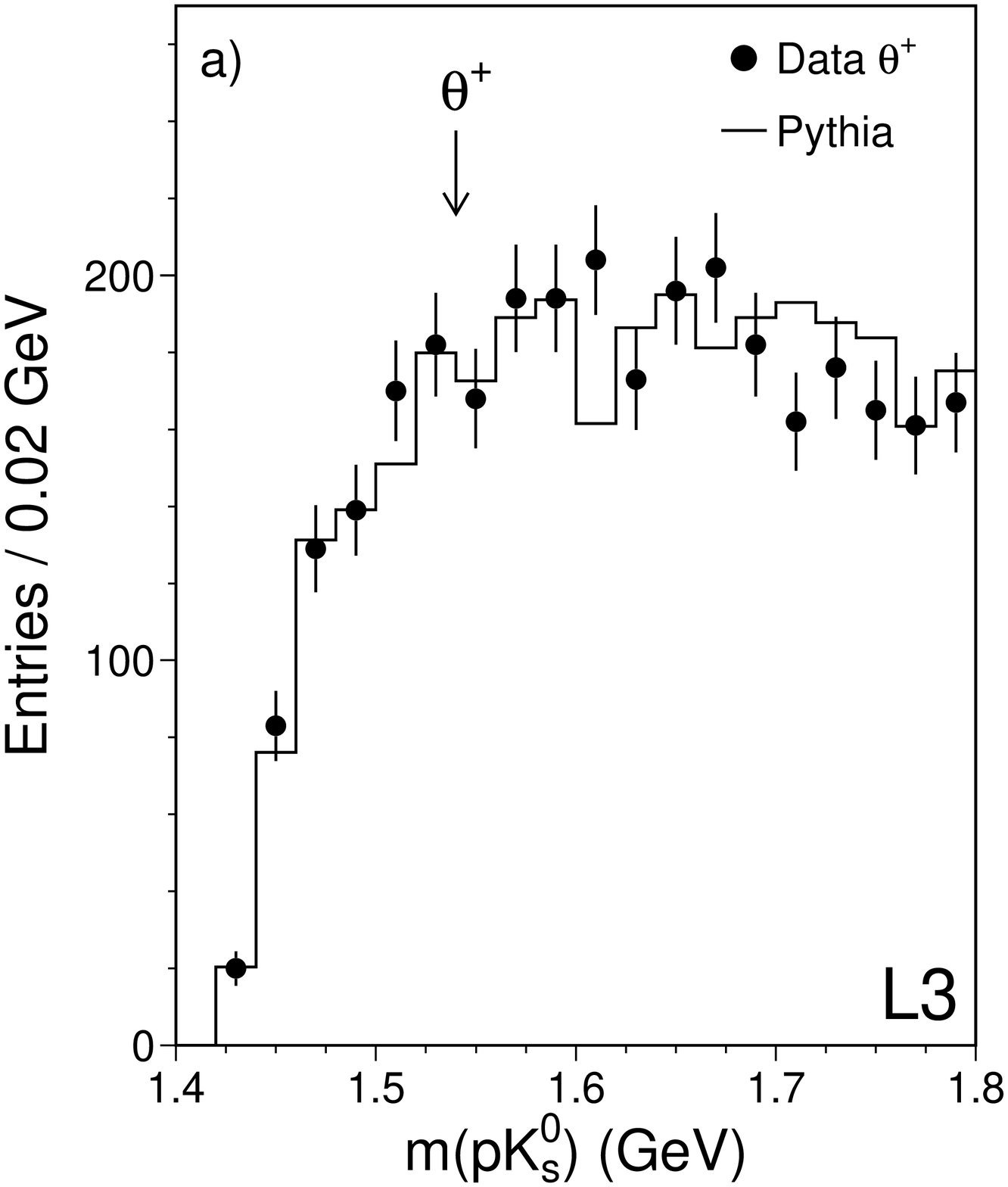,width=7.5cm}
\epsfig{file=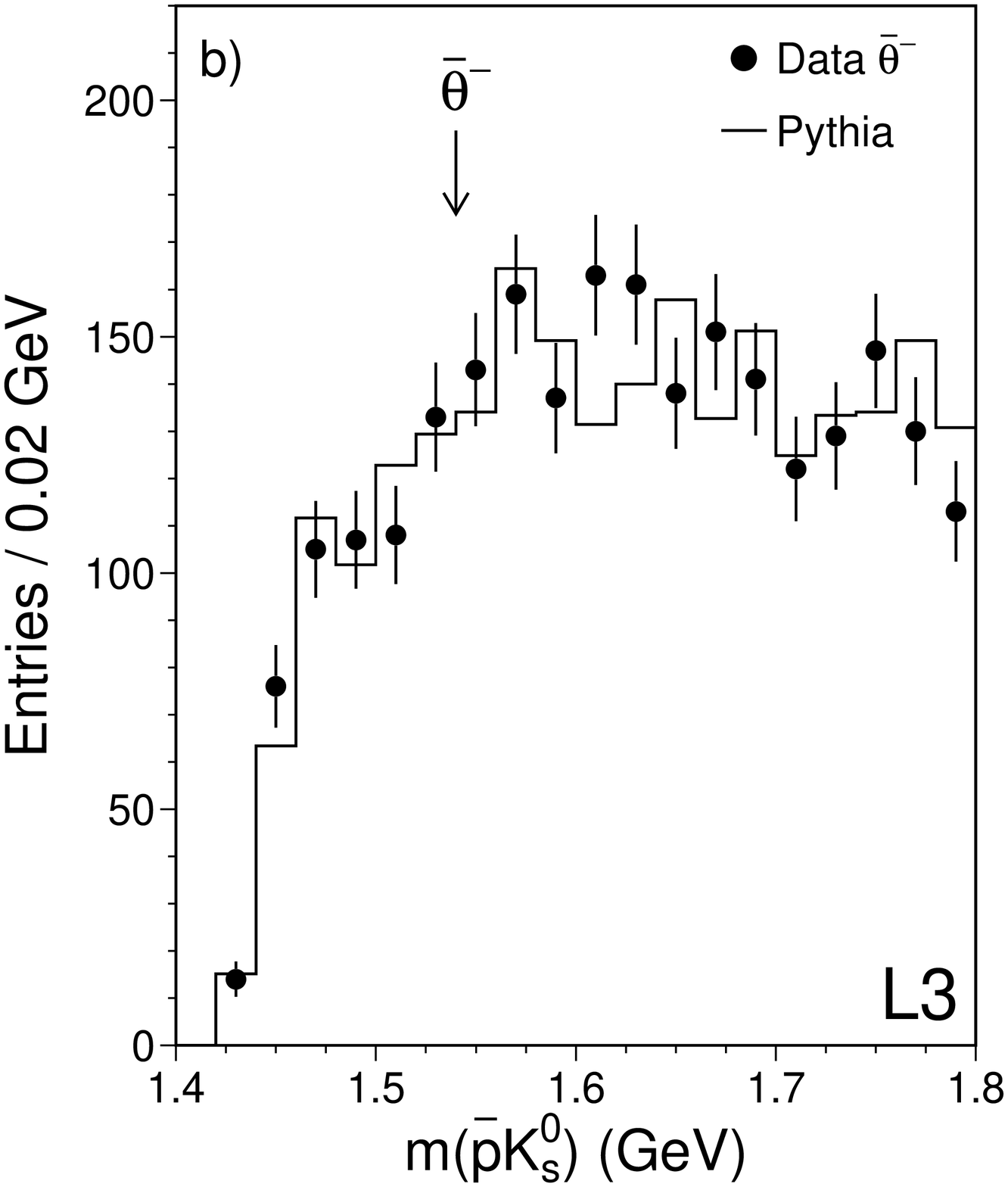,width=7.5cm}
\end{center}
\caption{The mass of the $\rm p \kos$ system, $\mpko$,  for a) protons and b) antiprotons 
together with the predictions of the PYTHIA Monte Carlo. The arrow indicates the position 
of the expected $\theta^+$ signal, around $1.54 \GeV$.}
\label{massthe}
\end{figure}

The resulting $\rm p \kos$ and $\rm \overline{p} \kos$ mass spectra, $m(\rm p \kos)$, 
are shown in Figure \ref{massthe}a and \ref{massthe}b respectively. Since no 
distinct structure due to the excess of protons is observed, the two spectra 
are combined in the following.

\subsection{Results}

\begin{figure}
\begin{center}
\epsfig{file=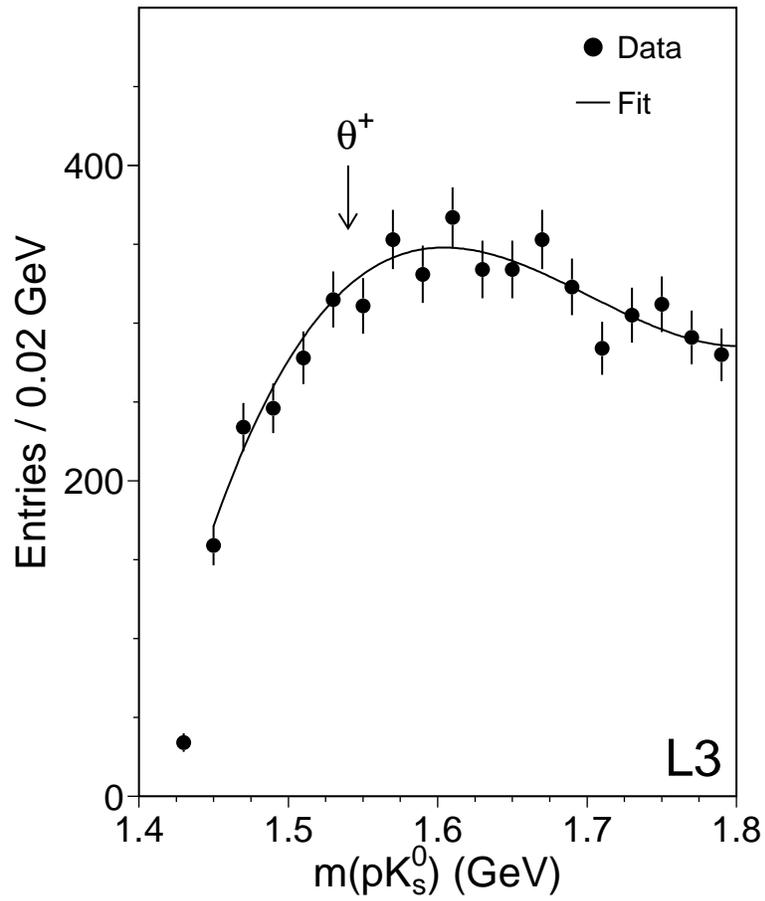,width=10cm} 
\end{center}
\caption{The mass of the $\rm p \kos$ system, $\mpko$, for protons and
antiprotons combined together with the fit result. The arrow indicates the position of the 
expected $\theta^+$ signal, around $1.54 \GeV$.}
\label{pentafit}
\end{figure}

An upper limit for the number of $\epem \ra \epem \theta^+ \rm X$ events 
at 95\% confidence level is derived with a fit to the $m{(\rm p \kos)}$ 
mass spectrum. To minimize the effect of the binning and the choice of the fitting 
interval, an unbinned maximum likelihood fit is performed in the range 
$1.45 \GeV  < \wgg < 1.8 \GeV$. The likelihood function is written as:
$${\cal L} = p \cdot g(m_{(\rm p \kos)}) + (1-p) \cdot b(m_{(\rm p \kos)}) $$
where $p$ denotes the fraction of signal events inside the fitted region. 
The background, $b(m_{(\rm p \kos)})$, is parametrized by a fourth-order polynomial and the signal 
by a Gaussian distribution, $g(m_{(\rm p \kos)})$. The resolution of the expected signal 
is determined from Monte Carlo to be $14.0 \pm 0.6 \MeV$. The 
resulting fit is displayed in Figure \ref{pentafit} and yields a fraction of 
signal events compatible with zero: $p=-0.006 \pm 0.007$. Restricting the 
measurement to the physical region $p \geq 0$, an upper limit of 59.3 events 
at 95\% confidence level is derived.

The $\theta^+$ selection efficiency is estimated to be 1.0\%, including a  branching fraction 
$\kos \ra \pi^+ \pi^-$ of 69\% \cite{pdg}. The branching ratio $\theta^+ \ra \rm p \kos$ is set 
to 1/4, accounting for competition with both the $\rm K^+ n$ and $\rm K^0_L p$ channels. 
The overall trigger efficiency is measured to be 83\%.

The mean number of $\theta^+$ pentaquark per hadronic two-photon event extrapolated 
to the full phase-space for $\wgg > 5 \GeV$ is found to be less than 
$4.0 \times 10^{-3}$ at 95\% confidence level. The systematic 
uncertainty arising from the selection procedure is evaluated by varying the cuts and 
repeating the fitting procedure. The contribution of the $\kos$ reconstruction 
is estimated to be 9\% whereas an uncertainty of 4\% is associated with proton 
identification. The uncertainty due to $\theta^+$ modelling, determined 
by using the $p_t$ and pseudo-rapidity distributions of different baryons in 
the $\theta^+$ generation, is found to be 9\%. 
The effect of a non-isotropic $\theta^+ \ra \rm p \kos$ decay on the efficiency 
is estimated using an angular distribution of the form $1-\cos^2\alpha$, where 
$\alpha$ is the angle of the proton in the $\theta^+$ centre-of-mass system. The associated 
systematic uncertainty is found to be 6\%. Limited Monte Carlo statistics 
introduces an additional uncertainty of 5\%. The sum in quadrature of these 
contributions yields a total systematic uncertainty of 16\%. Including this uncertainty 
in the determination of the number of $\theta^+$ pentaquark per two-photon 
hadronic event gives
$$ \langle n_{\theta^+} \rangle  < 4.7 \times 10^{-3}$$ 
at 95\% confidence level. This result is about four times greater than 
the mean $\Xi^*(1530)$ multiplicity: a baryon with almost the same mass.
This difference is mainly due to the stringent cuts 
applied to proton selection, resulting in a smaller selection efficiency than that of 
the $\Xi^*(1530)$ baryon, as well as the low branching ratio $\theta^+ \ra \rm p \kos$, 
considered as half of that of the decay $\Xi^*(1530) \ra \Lambda \pi^- \pi^+ \ra \rm p 
\pi^- \pi^- \pi^+$.

Using the two-photon cross-section for  $\wgg > 5 \GeV$, $\sigma_{\gamma \gamma} = 397 \; \rm nb$ \cite{l3tot}, a 95\% C.L. upper 
limit on the cross section $\gamma \gamma \ra \theta^+ X$ of 1.8 nb 
is obtained. This result is comparable to the upper limit obtained by 
the photoproduction experiment in the reaction $\gamma \rm p \ra 
\overline{\rm K}\,\!^0 \theta^+$ \cite{pentasearchnew}.

As a cross check, several subsets of the selected $\rm p \kos$ samples were 
investigated using tighter selection criteria. The $\kos$ selection interval 
was reduced to $\pm 10 \MeV$, the DCA of proton and antiprotons candidates was 
required to be less than 3 mm or secondary vertices compatible with the $\Lambda$ 
mass hypothesis were rejected. Since strangeness is a conserved quantity in strong 
interactions, the presence of an additional kaon was also required. No significant 
$\theta^+$ signal was observed in any of these samples.

\section{Conclusion}

The production of $\Lambda$, $\Xi^-$ and $\Xi^*(1530)$ baryons in two-photon 
collisions is studied in the range $0.4 \GeV < p_t <2.5\GeV$, $|\eta|<1.2$ 
and $\wgg > 5 \GeV$. The shape of the differential cross section for $\Lambda$ 
and $\Xi^-$ production is relatively 
well reproduced by the PYTHIA and PHOJET Monte Carlo programs using parameters tuned at 
$\sqrt{s}=m_Z$ although a better agreement can be obtained for $\Lambda$ by increasing the 
width in the gaussian $p_t$ distribution for primary hadrons.
The mean numbers of $\Lambda$, $\Xi^-$ and $\Xi^*(1530)$ baryons per hadronic 
two-photon event are found to be slightly below the PYTHIA and 
PHOJET predictions but overall agreement with 
the thermodynamical model is observed. 
The comparison between measurements obtained in two-photon events and 
$\epem$ annihilation processes provides evidence for the universality of 
fragmentation functions in both reactions.

Finally, a search for the pentaquark $\theta^+(1540)$ through the decay 
$\theta^+ \ra \rm p \kos$ for $\wgg > 5 \GeV$ is presented. No evidence for $\theta^+$ 
production is found and a 95\% confidence level upper limit on the mean number of 
$\theta^+$ per two-photon hadronic event at a level of four times the observed rate 
for the $\Xi^*(1530)$ baryon is derived.



\newpage
\input namelist308.tex
\newpage

\newpage


\begin{table}
\begin{center}
\begin{tabular}{|c|c|c|c|c|r@{~$\pm$~}r@{~$\pm$~}r|r@{~$\pm$~}r@{~$\pm$~}r|}
        \hline
$p_t$  & $\langle p_t \rangle$ & $\langle p_t^2 \rangle$ & Number of &Efficiency & \multicolumn{3}{|c|}{${\mathrm d}\sigma / {\mathrm d}p_t$} & \multicolumn{3}{|c|}{${\mathrm d}\sigma / {\mathrm d}p_t^2$} \\
(\GeV) &        (\GeV)         &   ($\GeV^2$)            & baryons   &(\%)       & \multicolumn{3}{|c|}{$\rm (pb/\GeV)$}                      & \multicolumn{3}{|c|}{$\rm (pb/\GeV^2)$}  \\ \hline
\multicolumn{11}{|c|}{$\Lambda$ baryon }\\ \hline
0.4$-$0.6 & 0.50 & 0.25  & $   3412 \pm 71$ & $10.2 \pm 0.1$ &  273.1& 5.7& 35.7&  273.1&    5.7&   35.7 \\
0.6$-$0.8 & 0.69 & 0.48  & $   4408 \pm 85$ & $13.7 \pm 0.1$ &  264.1& 5.1& 29.1&  188.7&    3.6&   20.8 \\
0.8$-$1.0 & 0.89 & 0.79  & $   3420 \pm 81$ & $15.3 \pm 0.2$ &  183.3& 4.4& 15.3&  101.9&    2.4&    8.5 \\
1.0$-$1.3 & 1.12 & 1.27  & $   3201 \pm 87$ & $16.8 \pm 0.2$ &  103.9& 2.8&  7.8&   45.2&    1.2&    3.4 \\
1.3$-$1.6 & 1.43 & 2.05  & $   1222 \pm 55$ & $17.7 \pm 0.4$ &   37.8& 1.7&  3.1&   13.0&    0.6&    1.1 \\
1.6$-$2.0 & 1.77 & 3.14  & $\phz578 \pm 41$ & $15.9 \pm 0.6$ &   15.0& 1.1&  1.4&    4.2&    0.3&    0.4 \\
2.0$-$2.5 & 2.21 & 4.92  & $\phz292 \pm 25$ & $17.2 \pm 1.3$ &    5.6& 0.5&  1.2&    1.2&    0.1&    0.3 \\
\hline
\multicolumn{11}{|c|}{$\Xi^-$ baryon }\\ \hline
0.4$-$0.7 & 0.55 & 0.31  & $\phz70 \pm  10$ & $\phz3.4 \pm 0.2$ & 11.3 &  1.7 &  1.7 &10.3 &1.5& 1.6 \\
0.7$-$1.0 & 0.83 & 0.70  & $   113 \pm  12$ & $\phz7.1 \pm 0.3$ &  8.7 &  1.2 &  1.2 & 5.1 &0.6& 0.7 \\
1.0$-$1.3 & 1.13 & 1.27  & $\phz83 \pm  12$ & $\phz9.7 \pm 0.8$ &  4.7 &  0.8 &  0.8 & 2.0 &0.3& 0.3 \\
1.3$-$2.5 & 1.67 & 2.88  & $\phz94 \pm  19$ & $\phz8.8 \pm 0.9$ &  1.5 &  0.3 &  0.3 & 0.4 &0.1& 0.1 \\
\hline
\multicolumn{11}{|c|}{$\Xi^*(1530)$ baryon }\\ \hline
0.4$-$2.5 & 0.82 & 0.84  & $\phz56 \pm  13$ & $\phz3.8 \pm 0.6$ &  1.4 &  0.3 &  0.5 & 0.5 &0.1& 0.2 \\
\hline

\end{tabular}
\end{center}
\caption{The average transverse momentum $\langle p_t \rangle$, the average transverse 
momentum squared $\langle p_t^2 \rangle$, the number of baryons estimated by the fits to 
the $\Lambda$, $\Xi^-$ and $\Xi^*(1530)$ mass spectra, the 
detection efficiency and the differential cross sections ${\mathrm d}\sigma / {\mathrm d}p_t$ 
and ${\mathrm d}\sigma /{\mathrm d}p_t^2$  for inclusive $\Lambda$, $\Xi^-$ and $\Xi^*(1530)$ production 
as a function of $p_t$ for $|\eta|<1.2$. The first uncertainty on the cross sections is 
statistical and the second systematic.}
\label{tabcrosspt}
\end{table}


\begin{table}
\begin{center}
\begin{tabular}{|c|c|c|c|r@{~$\pm$~}r@{~$\pm$~}r|}\hline
$|\eta|$  & $\langle |\eta| \rangle$ & Number of & Efficiency & \multicolumn{3}{|c|}{${\mathrm d}\sigma /{\mathrm d}|\eta|$} \\ 
          &                          & baryons   & (\%)       & \multicolumn{3}{|c|}{(pb)} \\ \hline
\multicolumn{7}{|c|}{$\Lambda$ baryon \hspace{1cm} $0.4\GeV < p_t < 1.0\GeV$}\\\hline

0.0$-$0.3 & 0.15&  $2953 \pm 72$  & $13.8 \pm 0.2$&   58.6&    1.4&    3.4\\
0.3$-$0.6 & 0.45&  $2742 \pm 63$  & $13.4 \pm 0.2$&   56.1&    1.3&    3.2\\
0.6$-$0.9 & 0.75&  $2904 \pm 70$  & $13.0 \pm 0.2$&   61.0&    1.5&    4.1\\
0.9$-$1.2 & 1.05&  $2774 \pm 89$  & $11.1 \pm 0.1$&   68.0&    2.2&    8.4\\
\hline
\multicolumn{7}{|c|}{$\Lambda$  baryon \hspace{1cm} $1.0\GeV < p_t < 2.5\GeV$}\\\hline
0.0$-$0.3 & 0.15&  $1458 \pm 60$  & $18.8 \pm 0.5$&   21.2&    0.9&    1.6\\
0.3$-$0.6 & 0.45&  $1411 \pm 62$  & $18.2 \pm 0.4$&   21.2&    0.9&    1.6\\
0.6$-$0.9 & 0.75&  $1480 \pm 62$  & $19.3 \pm 0.5$&   20.9&    0.9&    1.6\\
0.9$-$1.2 & 1.05&  $1007 \pm 62$  & $14.3 \pm 0.4$&   19.3&    1.2&    2.7\\
\hline
\multicolumn{7}{|c|}{$\Xi^-$ baryon \hspace{1cm} $0.4\GeV < p_t < 2.5\GeV$}\\\hline
0.0$-$0.4 & 0.20 & $ \phz110 \pm  14$ & $\phz6.7 \pm 0.3$ &    6.9& 0.9& 1.0\\
0.4$-$0.8 & 0.60 & $ \phz114 \pm  13$ & $\phz6.1 \pm 0.2$ &    8.0& 1.0& 1.0\\
0.8$-$1.2 & 1.01 & $ \phz109 \pm  15$ & $\phz5.5 \pm 0.2$ &    7.6& 1.1& 1.2\\
\hline
\end{tabular}
\end{center}
\caption{The average pseudo-rapidity $\langle |\eta| \rangle$, the number of baryons estimated 
by the fits to the $\Lambda$ and $\Xi^-$ mass spectra,
the detection efficiency and the differential
cross sections ${\mathrm d}\sigma / {\mathrm d}|\eta|$ for inclusive $\Lambda$ and $\Xi^-$ 
production as a function of $|\eta|$. The first
uncertainty on the cross sections is statistical and the second systematic.}
\label{tabcrosseta}
\end{table}


\begin{table}
\begin{center}
\begin{tabular}{|p{2.4cm}|c|c|c|} \cline{2-4}
\multicolumn{1}{c|}{} &Data & PYTHIA & PHOJET  \\\cline{2-4} 
\multicolumn{1}{c|}{} & \multicolumn{3}{|c|}{$0.4 \GeV < p_t < 2.5 \GeV$, $|\eta|<1.2$ and $\wgg>5 \GeV$} \\\hline
$\langle n_\Lambda \rangle $      & $(1.6\pm 0.1)\times10^{-2}$ & $(1.43\pm 0.01)\times 10^{-2}$ & $(1.80\pm 0.01)\times 10^{-2}$ \\
$\langle n_{\Xi^-} \rangle$       & $(7.6\pm 1.0)\times10^{-4}$ & $(8.61\pm 0.11)\times 10^{-4}$ & $(11.5\pm 0.10)\times 10^{-4}$ \\
$\langle n_{\Xi^*(1530)} \rangle$ & $(2.3\pm 1.0)\times10^{-4}$ & $(1.37\pm 0.04)\times 10^{-4}$ & $(1.91\pm 0.03)\times 10^{-4}$\\\hline
\end{tabular}
\vspace{1cm}

\begin{tabular}{|p{2.4cm}|c|c|c|} \cline{2-4}
\multicolumn{1}{c|}{} &Data & PYTHIA & PHOJET  \\\cline{2-4} 
\multicolumn{1}{c|}{} & \multicolumn{3}{|c|}{$\wgg>5 \GeV$} \\\hline
$\langle n_{\Lambda} \rangle $    & $(\phz7.6 \pm 0.8)\times10^{-2}$ & $(8.51\pm 0.01)\times 10^{-2}$ & $(1.11\pm 0.01)\times 10^{-1}$ \\
$\langle n_{\Xi^-} \rangle$       & $(\phz3.7 \pm 0.5)\times10^{-3}$ & $(5.17\pm 0.02)\times 10^{-3}$ & $(7.29\pm 0.09)\times 10^{-3}$ \\
$\langle n_{\Xi^*(1530)} \rangle$ & $(   11.8 \pm 5.2)\times10^{-4}$ & $(8.65\pm 0.09)\times 10^{-4}$ & $(12.4\pm 0.10)\times 10^{-3}$\\\hline
\end{tabular}
\caption{The mean number of $\Lambda$, $\Xi^-$ and $\Xi^*(1530)$ baryons 
per hadronic two-photon event for the measured $p_t$ and $|\eta|$ 
phase-space (top) and its extrapolation (bottom) to the full $p_t$ 
and $|\eta|$ range. The uncertainty on data includes 
both the statistical and systematic contributions whereas the uncertainty on the Monte 
Carlo predictions is statistical.}
\label{tabmult}
\end{center}
\end{table}


\begin{table}
\begin{center}
\begin{tabular}{|c|c| c| c |c | c| }
        \hline
$p_t$  & Selection  & Background       & Monte Carlo   & Monte Carlo & Total\\
(\GeV) & criteria   & subtraction      & statistics    & modelling   &      \\ \hline
%
\multicolumn{6}{|c|}{$\Lambda$ baryon  }\\\hline
%
0.4$-$0.6 &  \phz4.2&    12.1  & \phz0.9  & \phz1.4 &     13.1 \\
0.6$-$0.8 &  \phz4.2& \phz9.8  & \phz0.9  & \phz1.6 &     11.0 \\
0.8$-$1.0 &  \phz4.2& \phz6.6  & \phz1.2  & \phz1.9 &  \phz8.4 \\
1.0$-$1.3 &  \phz4.2& \phz5.3  & \phz1.4  & \phz2.1 &  \phz7.5 \\
1.3$-$1.6 &  \phz4.2& \phz5.8  & \phz2.4  & \phz2.4 &  \phz8.2 \\
1.6$-$2.0 &  \phz4.2& \phz6.4  & \phz3.5  & \phz2.8 &  \phz9.1 \\
2.0$-$2.5 &  \phz4.2&    19.4  & \phz7.5  & \phz4.8 &     21.8 \\
\hline
\multicolumn{6}{|c|}{$\Xi$ baryon  }\\\hline
%
0.4$-$0.7 &    10.3 & \phz8.7  & \phz5.8  & \phz3.1 &     15.1 \\
0.7$-$1.0 &    10.3 & \phz7.5  & \phz4.8  & \phz2.4 &     14.0 \\
1.0$-$1.3 &    10.3 & \phz9.1  & \phz8.3  & \phz3.3 &     16.5 \\
1.3$-$2.5 &    10.3 &    12.3  &    10.4  & \phz6.3 &     19.5 \\
\hline
\multicolumn{6}{|c|}{$\Xi^*(1530)$ baryon }\\\hline
%
0.4$-$2.5 &    19.2 &   24.6   &     16.5 &    13.6 &     37.9  \\
\hline

\end{tabular}
\end{center}

\caption{Systematic uncertainties in percent on the cross section of the $\epem \ra \epem
\Lambda \rm X $, $\epem \ra \epem \Xi^- \rm X $ and $\epem \ra \epem \Xi^*(1530) \rm X$ 
processes due to selection criteria, background subtraction, limited Monte Carlo statistics, 
Monte Carlo modelling and trigger efficiency as a function of $p_t$ for $|\eta|<1.2$. 
The total systematic uncertainty, taken as the quadratic sum of the different
contributions, includes a constant contribution of 2\% due to the determination 
of the trigger efficiency.}
\label{tabsyst}
\end{table}

\clearpage
\newpage


\end{document}

%% file: namelist308.tex
\typeout{   }     
\typeout{Using author list for paper 287 -  }
\typeout{$Modified: Jul 15 2001 by smele $}
\typeout{!!!!  This should only be used with document option a4p!!!!}
\typeout{   }
%
%
%
%
%
%

\newcount\tutecount  \tutecount=0
\def\tutenum#1{\global\advance\tutecount by 1 \xdef#1{\the\tutecount}}
\def\tute#1{$^{#1}$}
\tutenum\aachen            
\tutenum\nikhef            
\tutenum\mich              
\tutenum\lapp              
\tutenum\basel             
\tutenum\lsu               
\tutenum\beijing           
\tutenum\bologna           
\tutenum\tata              
\tutenum\ne                
\tutenum\bucharest         
\tutenum\budapest          
\tutenum\mit               
\tutenum\panjab            
\tutenum\debrecen          
\tutenum\dublin            
\tutenum\florence          
\tutenum\cern              
\tutenum\wl                
\tutenum\geneva            
\tutenum\hamburg           
\tutenum\hefei             
\tutenum\lausanne          
\tutenum\lyon              
\tutenum\madrid            
\tutenum\florida           
\tutenum\milan             
\tutenum\moscow            
\tutenum\naples            
\tutenum\cyprus            
\tutenum\nymegen           
\tutenum\caltech           
\tutenum\perugia           
\tutenum\peters            
\tutenum\cmu               
\tutenum\potenza           
\tutenum\prince            
\tutenum\riverside         
\tutenum\rome              
\tutenum\salerno           
\tutenum\ucsd              
\tutenum\sofia             
\tutenum\korea             
\tutenum\taiwan            
\tutenum\tsinghua          
\tutenum\purdue            
\tutenum\psinst            
\tutenum\zeuthen           
\tutenum\eth               

{
\parskip=0pt
\noindent
{\bf The L3 Collaboration:}
\ifx\selectfont\undefined
 \baselineskip=10.8pt
 \baselineskip\baselinestretch\baselineskip
 \normalbaselineskip\baselineskip
 \ixpt
\else
 \fontsize{9}{10.8pt}\selectfont
\fi
\medskip
\tolerance=10000
\hbadness=5000
\raggedright
\hsize=162truemm\hoffset=0mm
\def\r{\rlap,}
\noindent

P.Achard\r\tute\geneva\ 
O.Adriani\r\tute{\florence}\ 
M.Aguilar-Benitez\r\tute\madrid\ 
J.Alcaraz\r\tute{\madrid}\ 
G.Alemanni\r\tute\lausanne\
J.Allaby\r\tute\cern\
A.Aloisio\r\tute\naples\ 
M.G.Alviggi\r\tute\naples\
H.Anderhub\r\tute\eth\ 
V.P.Andreev\r\tute{\lsu,\peters}\
% ...
\newpage
\begin{list}{A}{\itemsep=0pt plus 0pt minus 0pt\parsep=0pt plus 0pt minus 0pt
                \topsep=0pt plus 0pt minus 0pt}
\item[\aachen]
 III. Physikalisches Institut, RWTH, D-52056 Aachen, Germany$^{\S}$
\item[\nikhef] National Institute for High Energy Physics, NIKHEF, 
     and University of Amsterdam, NL-1009 DB Amsterdam, The Netherlands
\item[\mich] University of Michigan, Ann Arbor, MI 48109, USA
\item[\lapp] Laboratoire d'Annecy-le-Vieux de Physique des Particules, 
     LAPP,IN2P3-CNRS, BP 110, F-74941 Annecy-le-Vieux CEDEX, France
\item[\basel] Institute of Physics, University of Basel, CH-4056 Basel,
     Switzerland
\item[\lsu] Louisiana State University, Baton Rouge, LA 70803, USA
\item[\beijing] Institute of High Energy Physics, IHEP, 
  100039 Beijing, China$^{\triangle}$ 
\item[\bologna] University of Bologna and INFN-Sezione di Bologna, 
     I-40126 Bologna, Italy
\item[\tata] Tata Institute of Fundamental Research, Mumbai (Bombay) 400 005, India
\item[\ne] Northeastern University, Boston, MA 02115, USA
\item[\bucharest] Institute of Atomic Physics and University of Bucharest,
     R-76900 Bucharest, Romania
\item[\budapest] Central Research Institute for Physics of the 
     Hungarian Academy of Sciences, H-1525 Budapest 114, Hungary$^{\ddag}$
\item[\mit] Massachusetts Institute of Technology, Cambridge, MA 02139, USA
\item[\panjab] Panjab University, Chandigarh 160 014, India
\item[\debrecen] KLTE-ATOMKI, H-4010 Debrecen, Hungary$^\P$
\item[\dublin] UCD School of Physics, University College Dublin, 
 Belfield, Dublin 4, Ireland
\item[\florence] INFN Sezione di Firenze and University of Florence, 
     I-50125 Florence, Italy
\item[\cern] European Laboratory for Particle Physics, CERN, 
     CH-1211 Geneva 23, Switzerland
\item[\wl] World Laboratory, FBLJA  Project, CH-1211 Geneva 23, Switzerland
\item[\geneva] University of Geneva, CH-1211 Geneva 4, Switzerland
\item[\hamburg] University of Hamburg, D-22761 Hamburg, Germany
\item[\hefei] Chinese University of Science and Technology, USTC,
      Hefei, Anhui 230 029, China$^{\triangle}$
\item[\lausanne] University of Lausanne, CH-1015 Lausanne, Switzerland
\item[\lyon] Institut de Physique Nucl\'eaire de Lyon, 
     IN2P3-CNRS,Universit\'e Claude Bernard, 
     F-69622 Villeurbanne, France
\item[\madrid] Centro de Investigaciones Energ{\'e}ticas, 
     Medioambientales y Tecnol\'ogicas, CIEMAT, E-28040 Madrid,
     Spain${\flat}$ 
\item[\florida] Florida Institute of Technology, Melbourne, FL 32901, USA
\item[\milan] INFN-Sezione di Milano, I-20133 Milan, Italy
\item[\moscow] Institute of Theoretical and Experimental Physics, ITEP, 
     Moscow, Russia
\item[\naples] INFN-Sezione di Napoli and University of Naples, 
     I-80125 Naples, Italy
\item[\cyprus] Department of Physics, University of Cyprus,
     Nicosia, Cyprus
\item[\nymegen] Radboud University and NIKHEF, 
     NL-6525 ED Nijmegen, The Netherlands
\item[\caltech] California Institute of Technology, Pasadena, CA 91125, USA
\item[\perugia] INFN-Sezione di Perugia and Universit\`a Degli 
     Studi di Perugia, I-06100 Perugia, Italy   
\item[\peters] Nuclear Physics Institute, St. Petersburg, Russia
\item[\cmu] Carnegie Mellon University, Pittsburgh, PA 15213, USA
\item[\potenza] INFN-Sezione di Napoli and University of Potenza, 
     I-85100 Potenza, Italy
\item[\prince] Princeton University, Princeton, NJ 08544, USA
\item[\riverside] University of Californa, Riverside, CA 92521, USA
\item[\rome] INFN-Sezione di Roma and University of Rome, ``La Sapienza",
     I-00185 Rome, Italy
\item[\salerno] University and INFN, Salerno, I-84100 Salerno, Italy
\item[\ucsd] University of California, San Diego, CA 92093, USA
\item[\sofia] Bulgarian Academy of Sciences, Central Lab.~of 
     Mechatronics and Instrumentation, BU-1113 Sofia, Bulgaria
\item[\korea]  The Center for High Energy Physics, 
     Kyungpook National University, 702-701 Taegu, Republic of Korea
\item[\taiwan] National Central University, Chung-Li, Taiwan, China
\item[\tsinghua] Department of Physics, National Tsing Hua University,
      Taiwan, China
\item[\purdue] Purdue University, West Lafayette, IN 47907, USA
\item[\psinst] Paul Scherrer Institut, PSI, CH-5232 Villigen, Switzerland
\item[\zeuthen] DESY, D-15738 Zeuthen, Germany
\item[\eth] Eidgen\"ossische Technische Hochschule, ETH Z\"urich,
     CH-8093 Z\"urich, Switzerland
\item[\S]  Supported by the German Bundesministerium 
        f\"ur Bildung, Wissenschaft, Forschung und Technologie.
\item[\ddag] Supported by the Hungarian OTKA fund under contract
numbers T019181, F023259 and T037350.
\item[\P] Also supported by the Hungarian OTKA fund under contract
  number T026178.
\item[$\flat$] Supported also by the Comisi\'on Interministerial de Ciencia y 
        Tecnolog{\'\i}a.
\item[$\sharp$] Also supported by CONICET and Universidad Nacional de La Plata,
        CC 67, 1900 La Plata, Argentina.
\item[$\triangle$] Supported by the National Natural Science
  Foundation of China.
\item[$\dagger$] Deceased.
\end{list}
}
\vfill
